\documentclass[preprint]{aastex}
\usepackage{graphicx}

\slugcomment{To appear in the Astronomical Journal}

\shorttitle{The Near-Ultraviolet Continuum of Late-Type Stars}
\shortauthors{Allende Prieto and Lambert}

\begin{document}


\title{The Near-Ultraviolet Continuum of  Late-Type Stars}


\author{Carlos Allende Prieto and David L. Lambert}
\affil{McDonald Observatory and Department of Astronomy, The University of Texas at Austin, RLM 15.308, Austin, Texas 78712-1083}


\begin{abstract}

Analyses of the near-ultraviolet continuum of late-type stars have led
to controversial results regarding the performance of state-of-the-art
model atmospheres. The release of the homogeneous IUE  final  archive
and the availability of the high-accuracy {\it Hipparcos} parallaxes
provide an opportunity to revisit this issue, as   accurate stellar
distances make it possible to compare observed absolute fluxes with
the predictions of model atmospheres.

The near-UV continuum is highly sensitive to $T_{\rm eff}$  and [Fe/H],
and once the gravity is constrained from the  parallax, these
parameters may be derived from the analysis of  low-dispersion {\it
long-wavelength} (2000--3000 \AA) IUE spectra for stars previously
studied by Alonso et al. (1996) using the Infrared Flux Method (IRFM).
A second comparison is carried out against the stars spectroscopically
investigated by Gratton et al. (1996).  It is shown that there is a
good agreement between $T_{\rm eff}$s obtained from
 the  IRFM and from the near-UV continuum, and a remarkable
correspondence between observed and synthetic fluxes for stars with
$4000 \le T_{\rm eff} \le 6000$ K of any metallicity and gravity. These
facts suggest that model atmospheres provide an adequate description of
the near-UV continuum forming region and that the opacities involved
are essentially understood.  For cooler stars, the results of the IRFM
are no longer reliable, as shown by Alonso et al., but the discrepancy
noticed for stars hotter than 6000 K may  reflect problems in the model
atmospheres and/or the opacities at these higher temperatures.

\end{abstract}




\section{Introduction}

The old problem of the missing opacity in the UV region of the solar
spectrum (Holweger 1970, Gustafsson et al. 1975) was claimed to be
solved by Kurucz (1992), who included millions of atomic and molecular
lines previously ignored in the computation of model atmospheres.
 Later,  Bell, Paltoglou \& Tripicco (1994) criticized that  solution,
 and the controversy  has been recently revived by Balachandran \& Bell
(1998) in connection with its relevance to the solar beryllium
abundance. In the mean time, Malagnini et al. (1992) and Morossi et al.
(1993) compared observations and   Kurucz's calculations for late-G and
early-K stars, and found that theory underpredicted the near-UV fluxes.
Very recently, other authors have  not found such inconsistencies in
the analysis of UV spectra for late-type metal-poor stars 
and also for O-B-A  stars
(Peterson 1999, Fitzpatrick \& Massa 1998, 1999a, 1999b). The situation is
confusing.  A reappraisal deserves to be made taking advantage of recent
revisions of   stellar near-UV fluxes measured by the IUE satellite and
the availability of  {\it Hipparcos} parallaxes (ESA 1997).

The continuum observed in the spectral region between 2500 and 3000
\AA\ is formed in the lower layers of the  photosphere for late-type
stars.  While  shorter wavelengths map higher atmospheric layers, this
spectral band is particularly important as a spectroscopic tool,
independent of the optical window,  to analyze the stellar
photosphere.  UV spectra are of relevance to the
determination of abundances of several astrophysically interesting elements
such as boron (Duncan et al. 1998) or neutron-capture elements such as
osmium, platinum, or lead  (Sneden et al. 1998). In a spectral region were spectral lines are highly crowded, a demostration  that observed fluxes match those predicted by the models used for the abundance analysis gives confidence in the derived abundances. In addition, it has been recognized in
the literature (Lanz et al. 1999) that good understanding of the near-UV 
spectrum of A$-$F stars is key for  dating intermediate-age stellar populations.

Accurate measurements of  stellar  fluxes in the ultraviolet are in principle
possible from outside Earth's atmosphere.  Absolute fluxes were first
measured through the long-wavelength cameras of the IUE satellite,
later the shuttle-borne WUPPE instrument, and finally through  GHRS,
and now its substitute STIS, onboard HST. The quality of the fluxes
measured by HST  is high, but  spectrographs onboard have mainly been
used  for high dispersion and therefore span a limited spectral
coverage.  The  long life of the IUE satellite provided an  extensive
dataset of low dispersion spectra, although even the 
recently released (NEWSIPS) version of the database has 
been found to include systematic effects (Massa \& Fitzpatrick 1998).  A
 newer version of the IUE Final Archive, named INES (IUE Newly
Extracted Spectra) has started to run at the time of writing this paper
(Rodr\'{\i}guez-Pascual et al. 1999).

Observations provide the flux at the Earth. Model atmospheres predict
the surface flux per unit area. Observation and prediction are related
by the stellar distance from Earth and the stellar radius. The absolute
magnitude calculated using the apparent visual magnitude, a bolometric
correction, and the {\it Hipparcos} parallax is combined with an
estimate of the effective temperature and theoretical evolutionary
tracks to derive the stellar radius. The radius and  the  {\it
Hipparcos} parallax make it possible to correct the observed flux for
dilution by the inverse-square law and so obtain the flux emitted from
the stellar surface.  Comparison with predicted fluxes is made for a
range in effective temperature and metallicity with the best fit to the
observed fluxes providing estimates of these two quantities. (Predicted
fluxes are weakly sensitive to surface gravity.) We compare these
estimates with those obtained by other techniques such as the InfraRed
Flux Method, and analysis of absorption lines in optical spectra.

\section{Observations}

IUE observations have been entirely reprocessed
 in a homogeneous fashion  with the set of procedures named NEWSIPS to
 produce the IUE Final Archive. This database, in particular the node
operated at  Villafranca  Satellite Tracking Station near
Madrid\footnote{http://iuearc.vilspa.esa.es/iuefab.html},  has been the
source of the spectra analyzed here. A newer version of the archive is
being released  through prototype servers (Rodr\'{\i}guez-Pascual et
al. 1999).

Several improvements are present in the 
 low-resolution NEWSIPS spectra employed here with
respect to  the older algorithms, such as a better weighted slit extraction
method, and a correction for the sensitivity degradation of the detectors
over the  life of the satellite and temperature variations.  An
improved procedure for obtaining the absolute flux calibrations was
also implemented. The reader is referred to the IUE NEWSIPS 
Information Manual  (Garhart et al. 1997) and references therein for  
more detailed information.

 When more than a single spectra was available for a given
star, they were combined and cleaned using the IUEDAC IDL Software
libraries\footnote{http://archive.stsci.edu/iue/iuedac.html} to produce
a single spectrum per star.  The effect of interstellar reddening 
 was considered negligible.

\section{The formation of the near-UV optical continuum and its
sensitivity to the basic atmospheric parameters}

We specifically refer to the  near-UV as the region  between 2000--3000
\AA. This spectral band is particularly interesting for the study of
stellar atmospheres, as it maps the deeper parts of the photosphere,
down below the region where the optical continuum is formed, but not as
deep as the continuum observed at 1.6 $\mu$m. A simple sketch of the
main hydrogenic opacities from  1 to 3 $\mu$m at a temperature of 5000
K and an electron pressure of 3 dyn cm$^2$  is shown in Fig. 1.
Hydrogen Rayleigh scattering, and even more importantly, but not
represented in Fig. 1, photoionization of carbon, silicon, aluminum,
magnesium, and iron produce a tremendous increase of the continuum
opacity for wavelengths shorter than about 2500 \AA\ (see Gray 1992 and
references therein), and  radiation is only able to escape from the
higher atmosphere.

\begin{figure}[ht!]
\centering
\includegraphics[width=9.cm,angle=90]{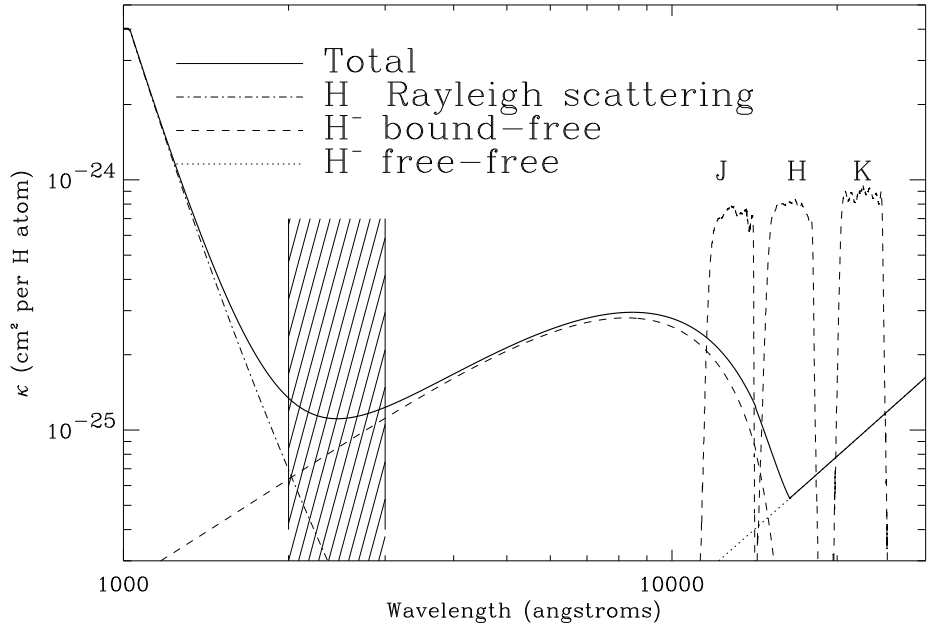}  
\protect\caption[]{Hydrogen Rayleigh scattering, and H$^{-}$ continuum opacity at 5000 K and 3 dyn cm$^{-2}$ in the near-UV, optical and near-IR. The
region between 2000 and 3000 \AA, in which we concentrate is highlighted,
and the wavelength coverage of the near-IR broad-band 
filters J, H, and K  is indicated.
\label{fig1}}
\end{figure}

\begin{figure}[ht!]
\centering
\includegraphics[width=9.cm,angle=90]{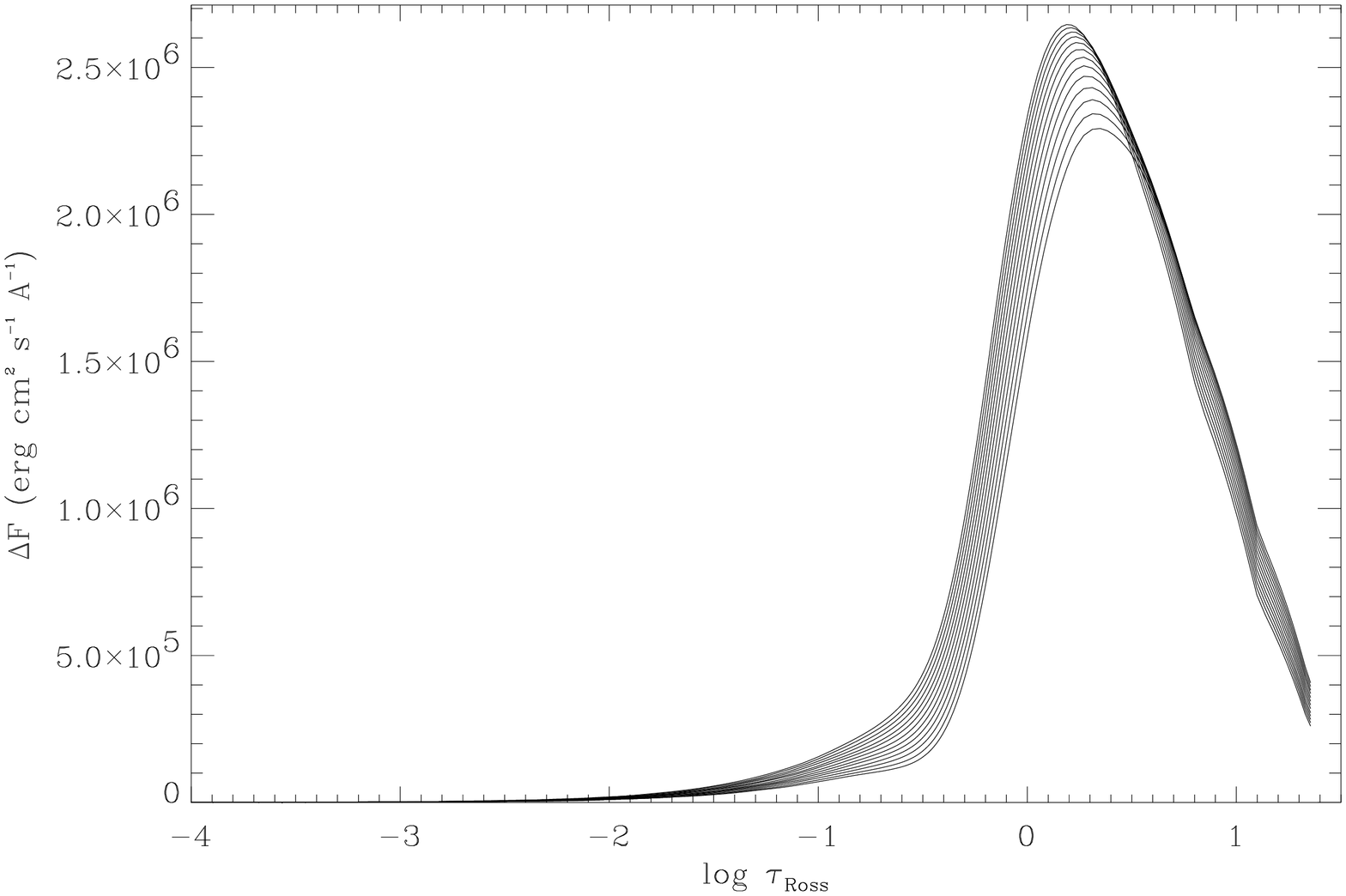}  
\protect\caption[]{Response function of the continuum at $2500-3000$ \AA\
 to the temperature, defined as the variation in the emerging flux produced
by a 10 \% change in the temperature at a given optical 
depth.
\label{fig2}}
\end{figure}

Between roughly  2000  and 2500 \AA\ and for solar abundances magnesium
photoionization dominates the continuum absorption , and the opacity is
larger, but of the same order of magnitude (yet uncertain) as the
H$^{-}$  in the optical and near-IR.  H$^{-}$ bound-free opacity is the
main contributor to the continuum opacity between 2500 and 3000 \AA.
  A quantitative measure of the formation depths of the continuum at
  those wavelengths for a solar-like photosphere  can be obtained
computing the response function to temperature. Figure 2 shows the
changes in the {\it true} continuum (not including line absorption) at
the stellar surface resulting from an increase of 10\% in temperature
at different atmospheric depths. The different lines correspond to
different wavelengths, and the longer the wavelength, the higher
(outer)  in the atmosphere the response function peaks. Therefore,
between roughly 2500 and 3000 \AA\, the  continuum
 is formed in deeper regions than the optical continuum, while  at
shorter wavelengths  the continuum covers higher layers.  Due to the
typical decrease of the flux towards shorter wavelengths
 in this region of the spectrum for late-type stars, and to the limited
signal-to-noise ratio in the IUE spectra, it is  the region between
2500 and 3000 \AA\ from where most of the information will be
retrieved.

We have made use of the flux distributions calculated by Kurucz, and
available at  CCP7\footnote{http://ccp7.dur.ac.uk} since 1993. The grid
includes models for different gravities ($\log g)$, effective
temperatures ($T_{\rm eff}$) and metal contents ([Fe/H]), while the
parameters in the mixing-length treatment of the convection are fixed,
as well as it is the microturbulence (2 km s$^{-1}$), and the abundance
ratio between different metals (solar-like mixture). For a given set of
($T_{\rm eff}$, $\log g$, [Fe/H]), we obtain the theoretical flux from
linear interpolation, and therefore using the information of the eight
nearest models available in the grid.

\begin{figure}[ht!]
\centering
\includegraphics[width=10.cm,angle=90]{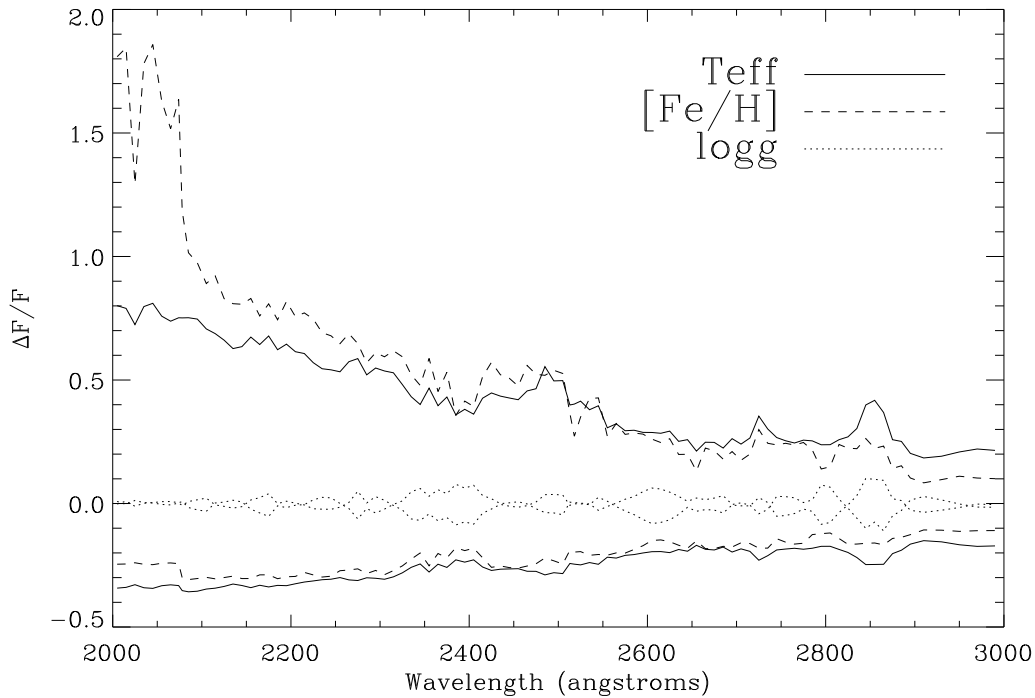}  
\protect\caption[]{Relative changes in the near-UV continuum in response to changes in the atmospheric parameters 
($T_{\rm eff}$, [Fe/H], $\log g$) for a  star like the Sun.
\label{fig3}}
\end{figure}

Neutral metals are  large contributors to  line absorption in the
near-UV. Therefore, temperature has three allied effects on the
emerging flux. Firstly, hotter temperatures increase the available
flux.  They also reduce the  importance of photodetachment absorption
of H$^{-}$, and so decrease the continuum absorption. Besides, the
diminished abundance of neutral metals reduces the line absorption. The
net effect is an important increment in the emerging flux.  The solid
lines in Fig. 3 show the result of a change of $\pm 100$ K in the
effective temperature for a solar model atmosphere. Changes in chemical
composition are important mainly for the neutral metal's contribution
to the line absorption, and this is demonstrated by the dashed lines,
which correspond to modifying in $\pm 0.2$ dex the logarithm of the
solar metal abundance. Gas pressure plays a minor role, as reflect the
dotted lines in Fig. 3, which correspond to changes in gravity of a
70\%. The effects of $T_{\rm eff}$ and the metal content are
significant, and both leave  characteristic signatures on the absolute
flux resulting from the different shape of the line and continuum
absorption. This will make it possible to estimate these two stellar
parameters from the observed absolute fluxes.   Fig. 3 shows
that the  changes in the slope of the observed spectrum induced by
variations of $T_{\rm eff}$ or [Fe/H] are  more subtle; it is much more
difficult to extract the information on the atmospheric physical
conditions from {\it relative} (not absolute) measurements of the
spectral energy distribution in these wavelengths, as  already demonstrated by Lanz et al. (1999).

\section{Near-UV fluxes as a tool to derive stellar parameters}

The modeling of late-type stellar spectra in the near-UV region
presents  all the same problems as any other spectral window. The
adequacy of the assumptions involved  in the construction of  model
atmospheres is  critical.  Line blanketing affects the  structure of
stellar  photospheres (Mihalas 1978), but as an extra difficulty, in
the near-UV the concentration of lines is so high as to give shape to
the overall energy distribution.

Early confrontation of UV fluxes predicted by classical model
atmospheres with observations (Holweger 1970, Gustafsson et al. 1975)
revealed inconsistencies, the predicted fluxes exceeding observations.
Later Kurucz (1992) claimed to have solved the problem by including
previously missing line absorption. Bell, Paltoglou \& Tripicco (1994)
presented evidence that Kurucz to had included  too many  lines. A
comparison at high dispersion in the regions 3400--3450 \AA\ and
4600--4650 \AA\  revealed that synthetic spectra based on Kurucz's
linelist  predicted stronger-than-observed absorption features.  Bell
et al.  (1994), and later Balachandran \& Bell (1998), suggested
missing contributors to the continuum absorption rather than line
blanketing as a possible explanation for the problem. It is unclear
whether   Kurucz's calculations experience that weakness in the
spectral window we  concentrate on (2000--3000 \AA), but while
comparison of synthetic spectra and observations of the Sun (or any
other single star)  will  leave room for  line absorption to mimic
missing continuum opacity, or viceversa, simultaneous comparison with a
number of stars of different temperatures, and in particular, chemical
compositions, will strongly limit that possibility.

\begin{figure}[ht!]
\centering
\includegraphics[width=9.cm,angle=90]{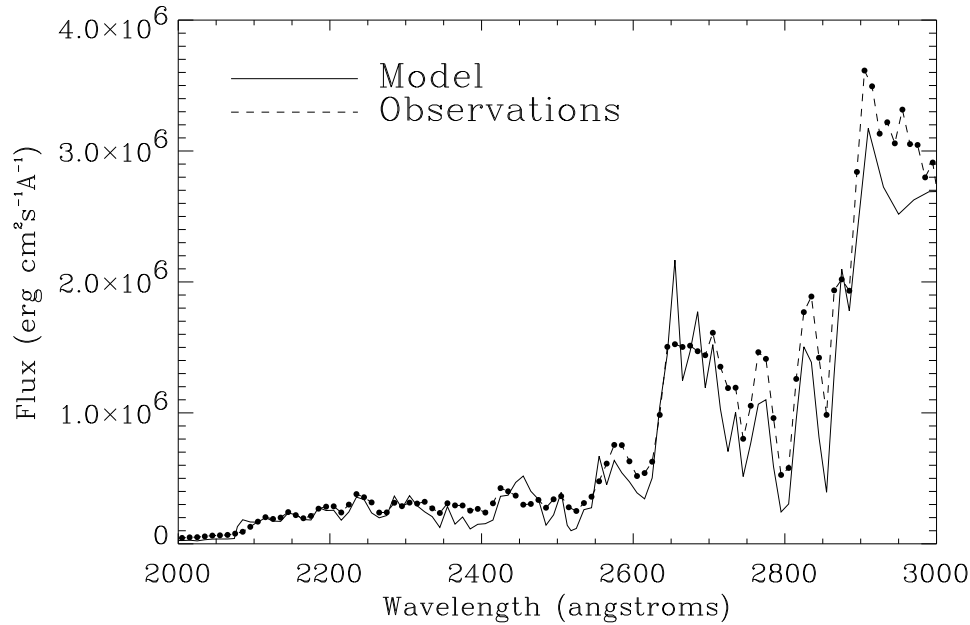}  
\protect\caption[]{Comparison between observed (dashed line; also dots) 
and predicted  (solid line) near-UV  fluxes at the surface of the Sun.
\label{fig4}}
\end{figure}

In any case, comparison with the Sun is a must. Colina, Bohlin \&
Castelli (1996)  compiled
an updated version of the available measurements of the solar flux
distribution. Fig. 4 shows fairly good agreement with the theoretical
predictions, in consistency with Kurucz's claims. The fluxes are compared at 
the solar surface.

The stellar parameters are known for no star with  such an extremely high
accuracy as for the Sun. However, semi-empirical methods to derive
$T_{\rm eff}$s have been applied to solar-metallicity stars. The
Infrared Flux Method (IRFM; Blackwell et al. 1991) is weakly dependent
on the model atmospheres, and in particular, the  line blanketing only
affects  the atmospheric structure, but not  the calculation of the
flux itself. The procedure's reliability has been tested
 with temperatures obtained from  measurements of angular diameters by
lunar occultation.

In the following sections we compare $T_{\rm eff}$s and [Fe/H]s for the
 stars with metallicities in the range $ -3.5 \le$ [Fe/H] $\le +0.5$
studied by Alonso et al. (1996) with those obtained from the comparison
of predicted and observed near-UV fluxes. The IRFM has been  applied by
Alonso et al.  to a large sample of late-type dwarfs and subgiants with
either spectroscopic or photometric estimates of the metallicity,
making it possible to constrain the second fundamental parameter that
influences the near-UV continuum.  Despite  the claimed weak-dependence
of the results on the choice of model atmosphere, it is interesting to mention
that the models employed in this study are similar, if not identical,
to those used by Kurucz to compute the predicted flux distributions
employed here. Gratton et al. (1996) made use of the published results
from the IRFM to  construct a reference frame of stars, and used it in
combination with other spectroscopic indicators, such as the iron
ionization balance,  to derive stellar parameters for a larger sample
of stars. Again, similar or identical model atmospheres are involved.

Our analysis   adopts the following scheme:

\begin{enumerate}
\item Estimates of the stellar mass ($M$), and bolometric correction ($BC$)  
are obtained following the same procedure as in Allende Prieto et al. (1999).
Briefly, the {\it Hipparcos} parallaxes ($p$) are used to transform visual $V$
magnitudes to absolute $M_V$ magnitudes. Depending on the metallicity,
an isochrone from the calculations by Bergbusch \& VandenBerg (1992) is
then used to  estimate  $M$ and $BC$,
interpolating in the $M_V-M$ and $M_V-BC$ relationships. Here it is assumed that
 stars with [Fe/H] $> - 0.47$ have an age of 9 $\times 10^9$ years, and those 
with [Fe/H] $< - 0.47$ are 12 $\times 10^9$ years old, although this is has
a negligible relevance (see Allende Prieto et al. 1999).

\item Using the initial estimates for the effective temperature from a
source (e.g. Alonso et al. 1996),  $T_{{\rm eff}}^0$, 
the gravities and radii are then obtained through the well-known expressions:
\begin{equation}
\label{logg}
{\rm log} \frac {g} {g_{\odot}} = {\rm log} \frac {M} {M_{\odot}}
  + 4{\rm log} \frac {T_{{\rm eff}}^0} {T_{{\rm eff},\odot}} +0.4 V
  +0.4{\rm BC} +2{\rm log}p +0.12, {  \rm and}
\end{equation}

\begin{equation}
\label{logr}
{\rm log} \frac {R} {R_{\odot}} = \frac{1}{2} \left({\rm log} \frac {g} {g_{\odot}} - {\rm log} \frac {M} {M_{\odot}}\right).
\end{equation}

\item The near-UV IUE spectra are compared with the synthetic spectra,
after converting the flux predicted at the stellar surface to Earth
using the nondimensional dilution factor $(pR)^2$, deriving the values
of $T_{\rm eff}$ and [Fe/H] that  minimize, in the least-square
sense, their differences. This is
performed using the Nelder-Mead  simplex method  for
multidimensional minimization of a function, as implemented by Press et
al. (1988), giving even weights to all wavelengths.

\item The gravity is then modified to be consistent with the new $T_{\rm eff}$:
\begin{equation}
\label{logg}
{\rm log} \frac {g} {g_{\odot}} =  {\rm log} \frac {g} {g_{\odot}} 
- 4{\rm log} \frac {T_{{\rm eff}}^0} {T_{{\rm eff},\odot}} 
+ 4{\rm log} \frac {T_{\rm eff}} {T_{{\rm eff},\odot}},  
\end{equation}

\noindent while variations in other magnitudes resulting from corrections in
[Fe/H] were found to be negligible.

\item Then, final values for $T_{\rm eff}$ and [Fe/H] are derived from a new
comparison between synthetic and observed spectra.
\end{enumerate}

The transfer of errors in  gravity  and  distance determined from the
{\it Hipparcos} parallax (see Allende Prieto et al. 1999) to errors in
the derived $T_{\rm eff}$ and [Fe/H] was estimated computing upper and
lower limits to the  dilution factor  $(pR)^2$,  and repeating the
minimization of the differences between observed and predicted fluxes.
The gravity is decreased and the dilution factor increased by the
estimated uncertainties to produce upper limits for $T_{\rm eff}$ and
lower limits for [Fe/H], and the signs of the increments are reversed
to obtain lower limits for $T_{\rm eff}$ and upper limits for [Fe/H].
This is generally appropriate, especially because errors in the flux
dilution factors  typically produce a much larger impact than those in
the gravity. In a very few cases, when the internal uncertainties are
particularly small, the rule of positive superindices (upper limits)
and negative subindices (lower limits) in the derived $T_{\rm eff}$s
and [Fe/H]s  shown in  Tables 1 and 2 is broken.  The use of the
Nelder-Mead simplex method to find the best fit to the observed spectra
is well justified, as for all extreme cases checked, a single minimum
was present, and the $\chi^2$ was found to vary smoothly with $T_{\rm
eff}$ and [Fe/H]. No changes were made in the original resolution of
observed or calculated fluxes, as they were similar enough for our
purposes.

\subsection{Comparison with the $T_{\rm eff}$s derived by Alonso et al. (1996) from the IRFM}

316 low dispersion spectra of 88 of the stars observed by {\it
Hipparcos} in the sample of Alonso et al. (1996) were obtained with the
low-dispersion long-wavelength cameras of IUE. The reddenings listed by
Alonso et al.  were taken into account to derive the gravities from the
{\it Hipparcos} parallaxes. Two stars (HR3427, HR8541) were discarded
as they were too hot ($T_{\rm eff} > 8000$ K) for the selected
isochrones.  Eleven more stars (G099-015, G119-052, G171-047, G231-019,
HD140283, HR1084, HR2943, HR4030, HR4623. HR509, HR937) were dropped as either the quality of their spectrum was extremely poor and/or the procedure to fit the spectrum failed (we recall that the interstellar extinction is being neglected).

\begin{figure}[ht!]
\centering
\includegraphics[width=5.cm,angle=90]{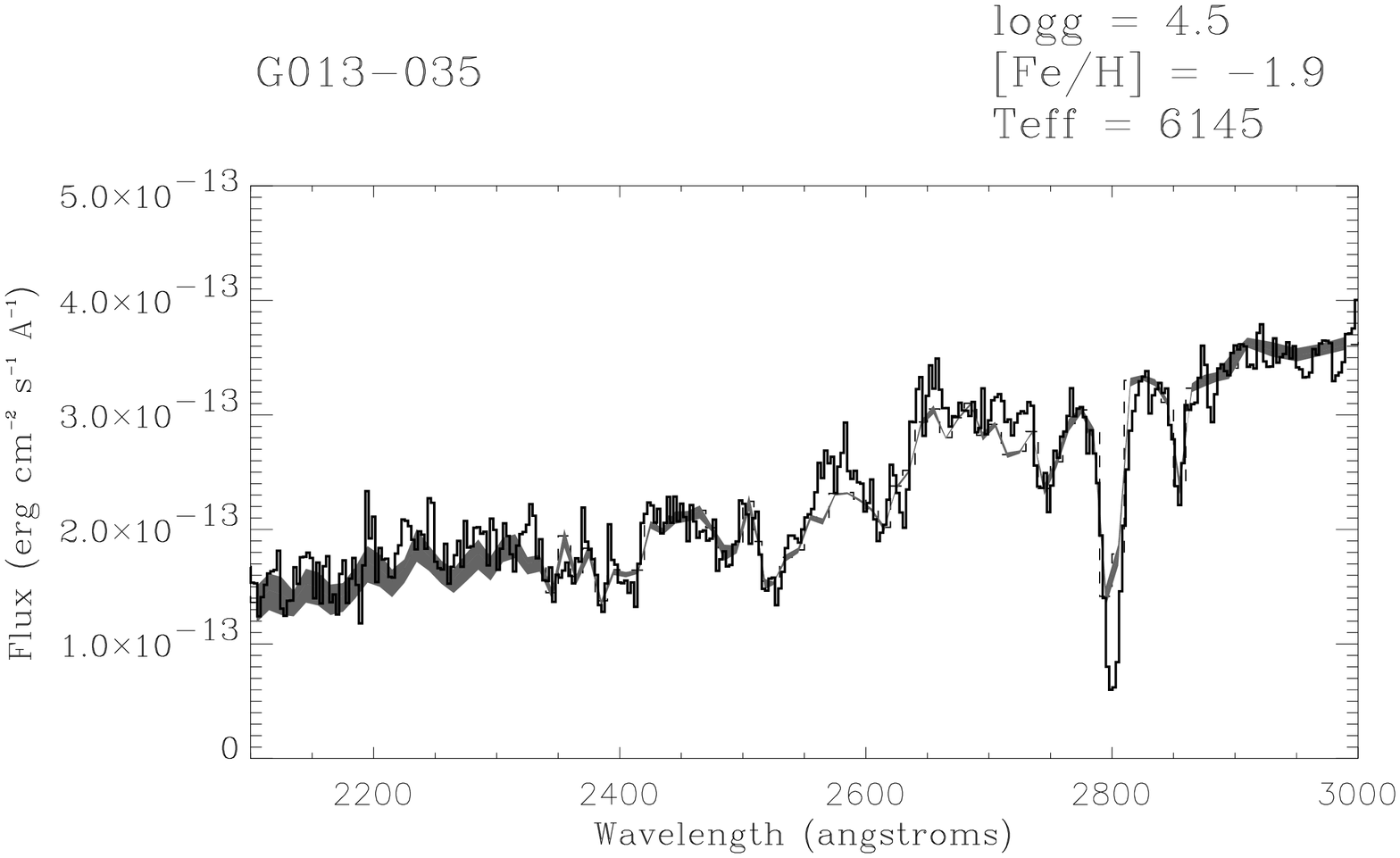}  
\includegraphics[width=5.cm,angle=90]{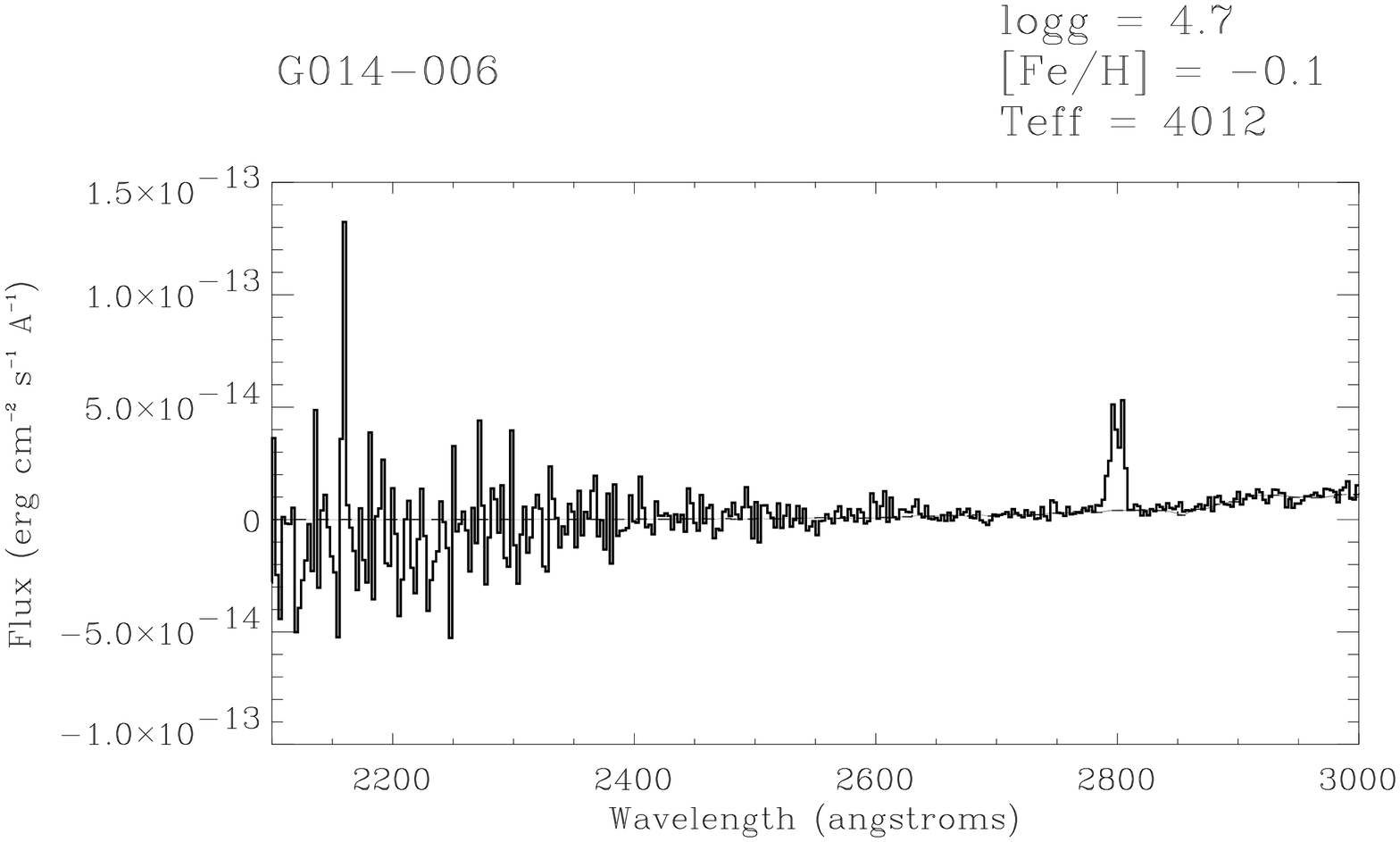}  
\includegraphics[width=5.cm,angle=90]{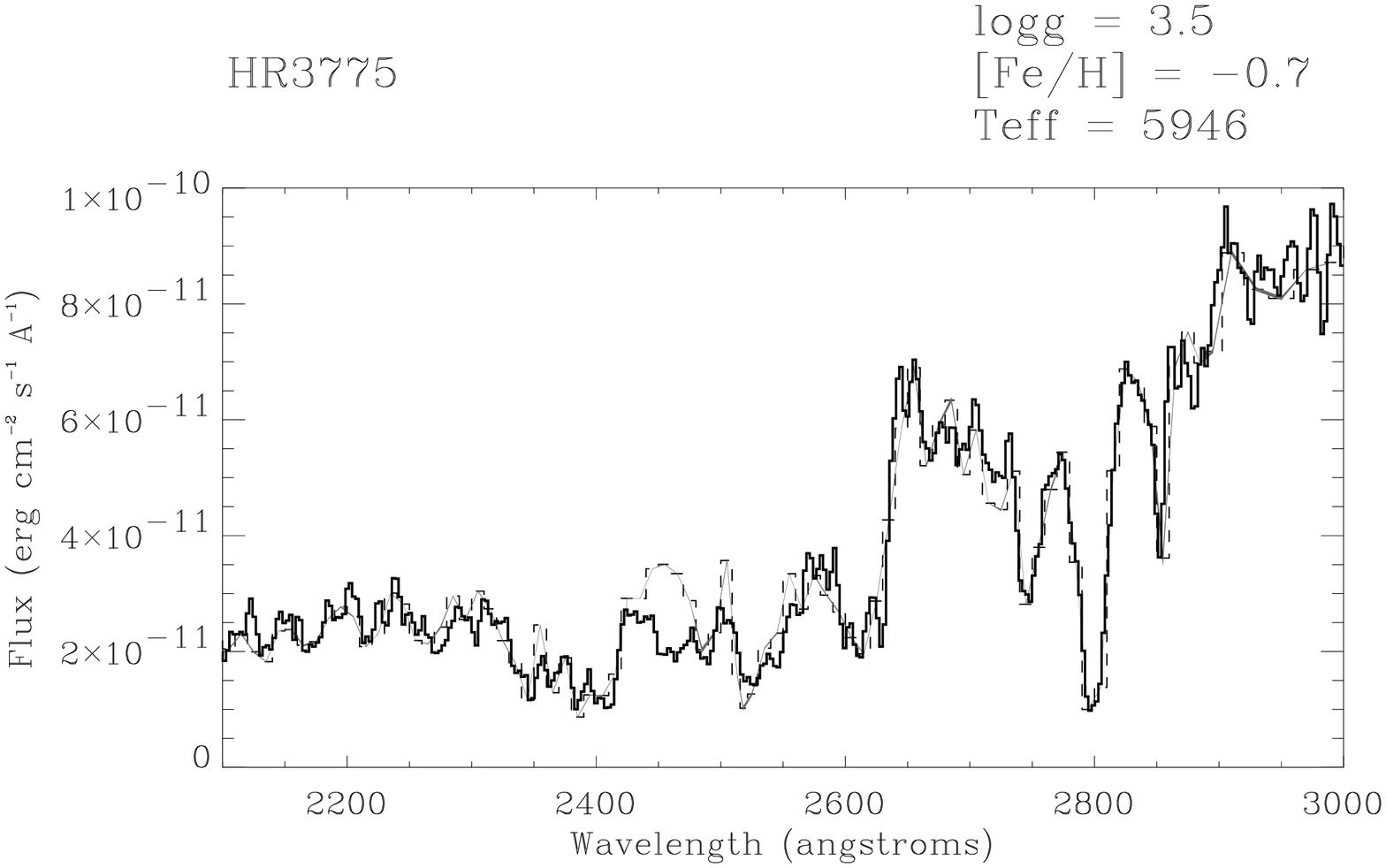}  
\includegraphics[width=5.cm,angle=90]{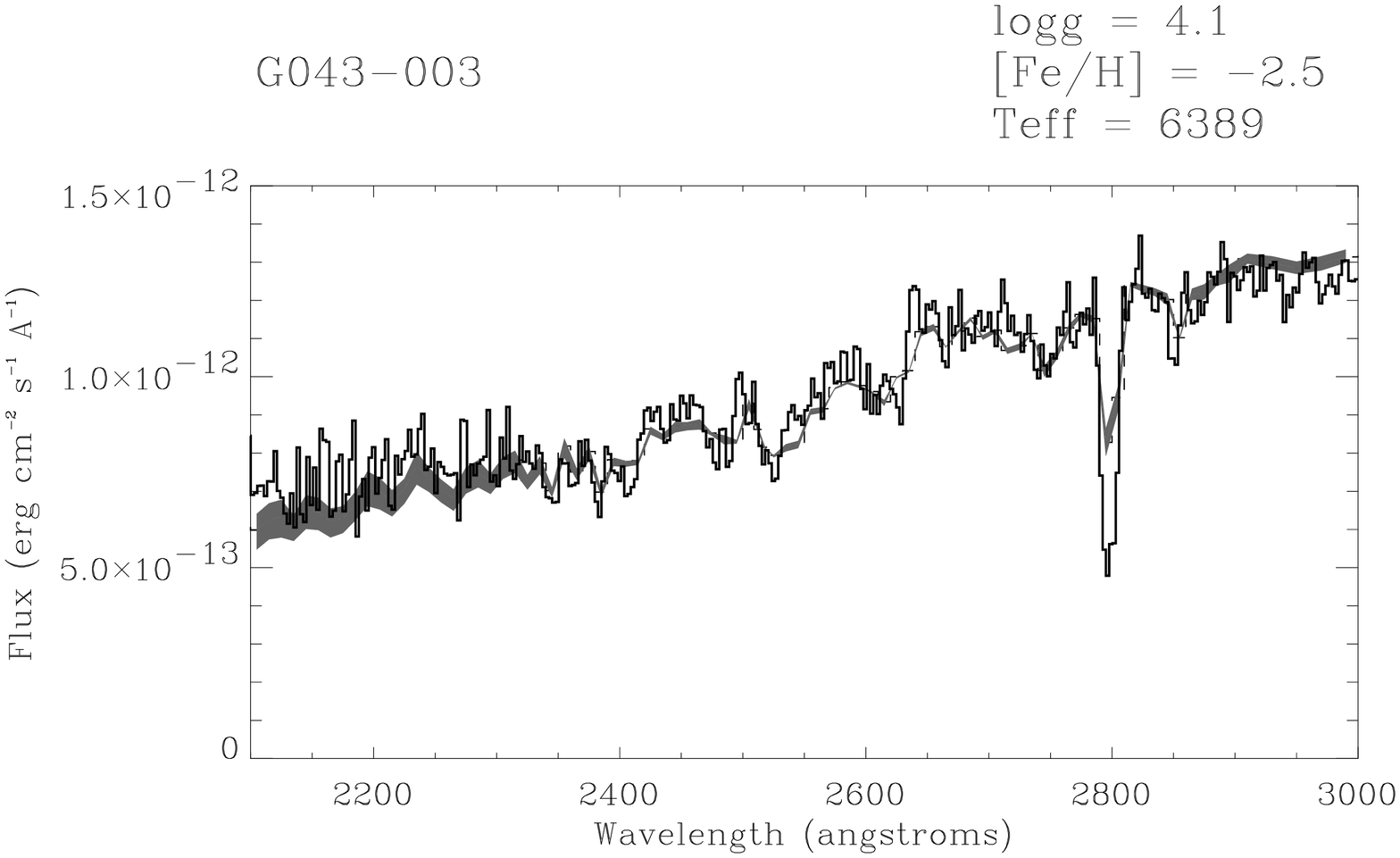}  
\includegraphics[width=5.cm,angle=90]{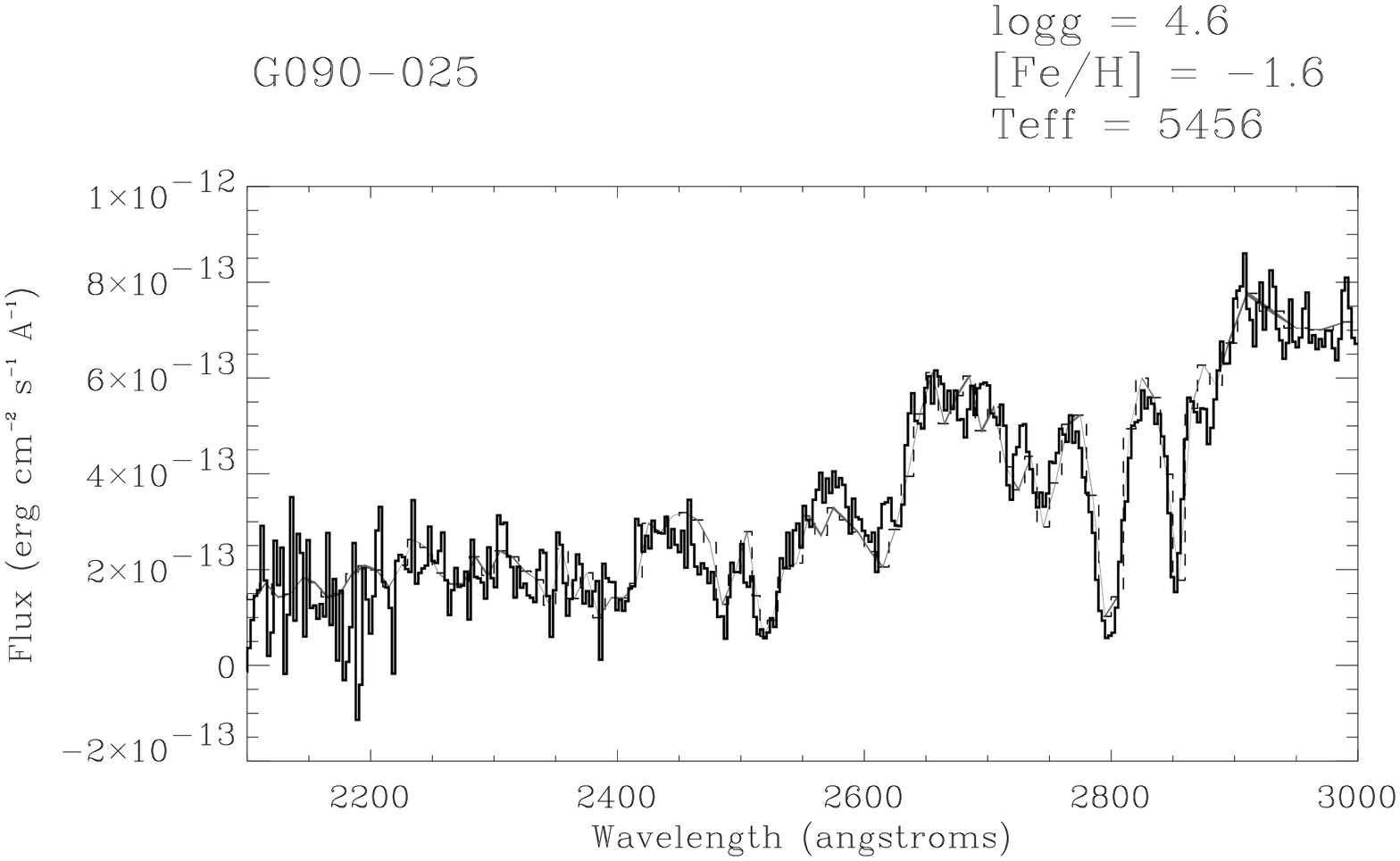}  
\includegraphics[width=5.cm,angle=90]{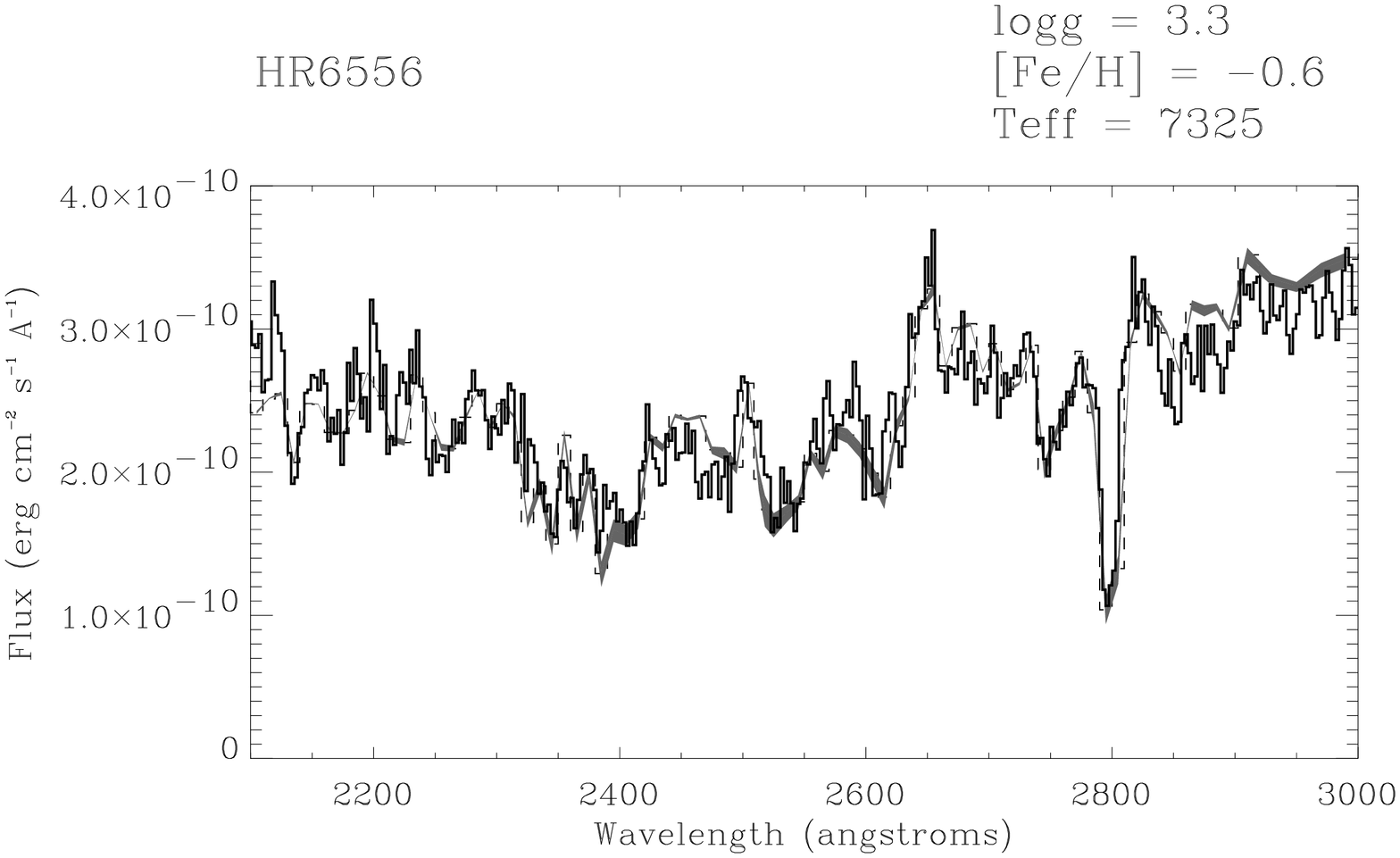}  
\protect\caption[]{Observed (thick solid line) and theoretical (shaded and broken lines) fluxes  at Earth for six of the
stars analyzed by Alonso et al. (1996). 
The thickness of the  shaded lines represents the range of 
possible fits resulting from uncertainties in  the derived gravity and 
dilution factor $(pR)^2$.
\label{fig5}}
\end{figure}

Fig. 5 displays several examples of the  comparison between theoretical
fluxes at  Earth (shaded and broken lines) and the IUE
observations (thick solid line).  The thickness of the shaded lines 
indicates  the different fits obtained when upper and lower limits of the 
errors in the flux dilution
factor  are taken into account and correspond to different values of
$T_{\rm eff}$ and [Fe/H]. The finally derived
stellar parameters for all the stars, and their lower and upper limits
are listed in Table 1. It is possible to find  a pair ($T_{\rm eff}$,
[Fe/H]) that reproduce the observed fluxes within the uncertainties;
the final match of the energy distribution is excellent. A strong
discrepancy is evident between  predicted and observed strength
of the Mg I resonance line at 2852 \AA\ in the spectra of metal-deficient stars. Magnesium is one of the 
so-called $\alpha$-elements, whose abundance ratio to iron is known
to be larger than solar in metal-poor stars, a fact not taken into account
in the construction of the model atmospheres and the calculation of the synthetic spectra used here.

The comparison of the IRFM effective temperatures published by Alonso
et al.  (1996) as the averaged values from the application of the
method in the J, H, and K broad bands with the values obtained from the
fit to the near-UV flux is shown in Fig. 6 (upper panel).
 The mean difference is only $-0.3$\%, and the standard deviation is
3\%.  However,  the level of agreement is not evenly distributed along
the temperature range. The standard deviation reduces  to
2\% for the stars with $4000 \le T_{\rm eff} \le 6200$ K.  

For stars cooler than 4000 K, molecular absorption plays a major role,
and it has been recognized many times that the models used here do not
include this absorption adequately.  Evidence for this is abundant in
the literature, and to mention an example particularly relevant to this
comparison, the  internal consistency found by Alonso et al. (1996)
among the $T_{\rm eff}$s derived from the different bands disappears
for stars cooler than 4000 K.  Besides, Alonso et al.  have shown that
the sensitivity to errors in the input quantities of the IRFM becomes
particularly enhanced for those stars.  For stars hotter than about
6500 K, neutral hydrogen photoionization makes an increasingly
important contribution to the continuum opacity in the optical, near IR,  and
near-UV. The fact that IRFM temperatures show high internal consistency
for stars with $6500 \le T_{\rm eff} \le 8500$ K but  do not agree with
those derived from fitting the near-UV continuum may reveal an
important inconsistency of the model atmospheres or errors in the UV
opacity at those temperatures. However, it is not possible  to rule out
other possibilities at this stage. For example, we have not explored
the influence of a change in the parameter(s) involved in the
mixing-length treatment of the convection, the microturbulence, the
binning of both the models and the observations, or the presence of
systematic errors in the flux calibration.

\begin{figure}[ht!]
\centering
\includegraphics[width=8.cm,angle=90]{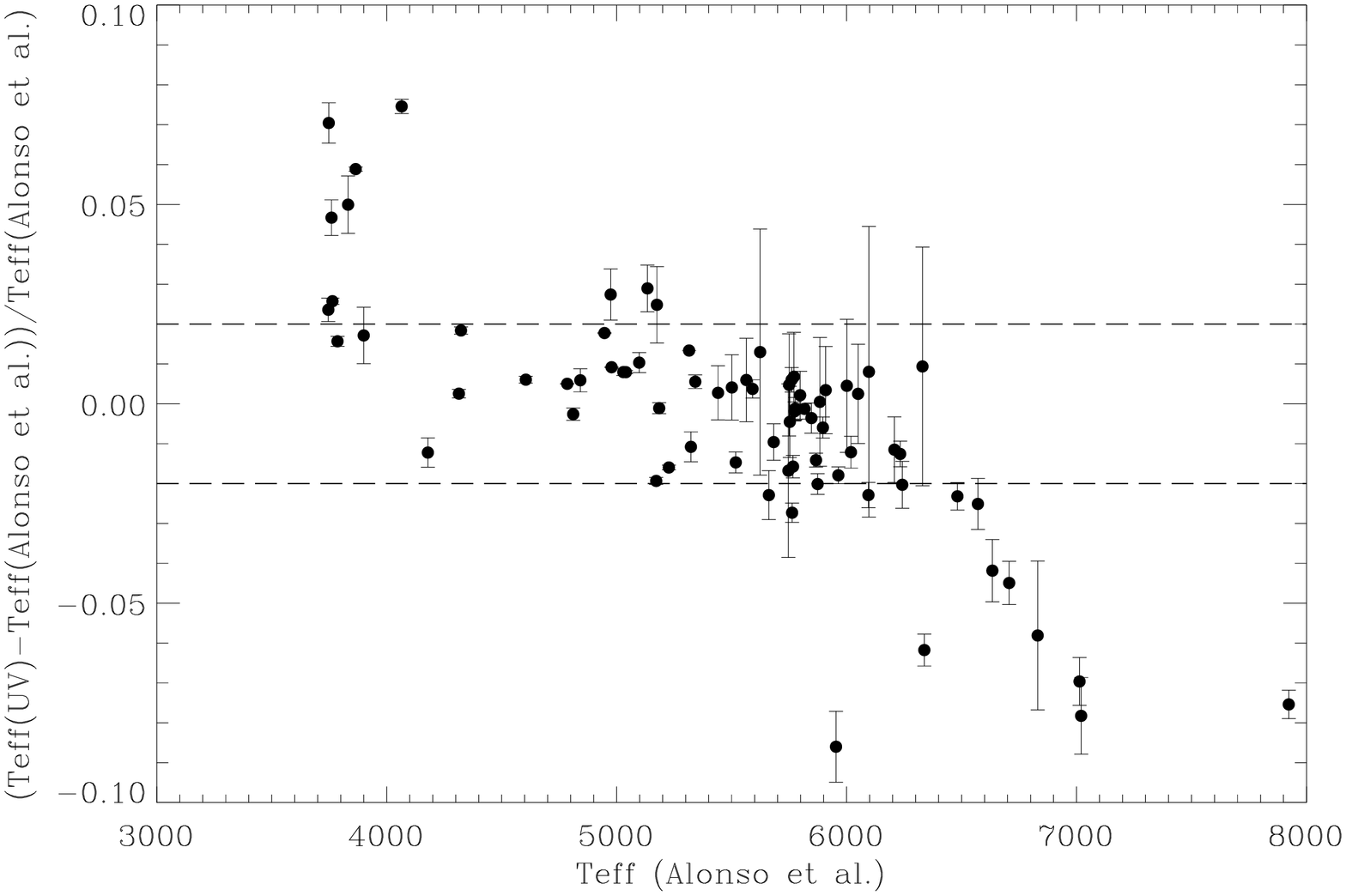}  
\includegraphics[width=8.cm,angle=90]{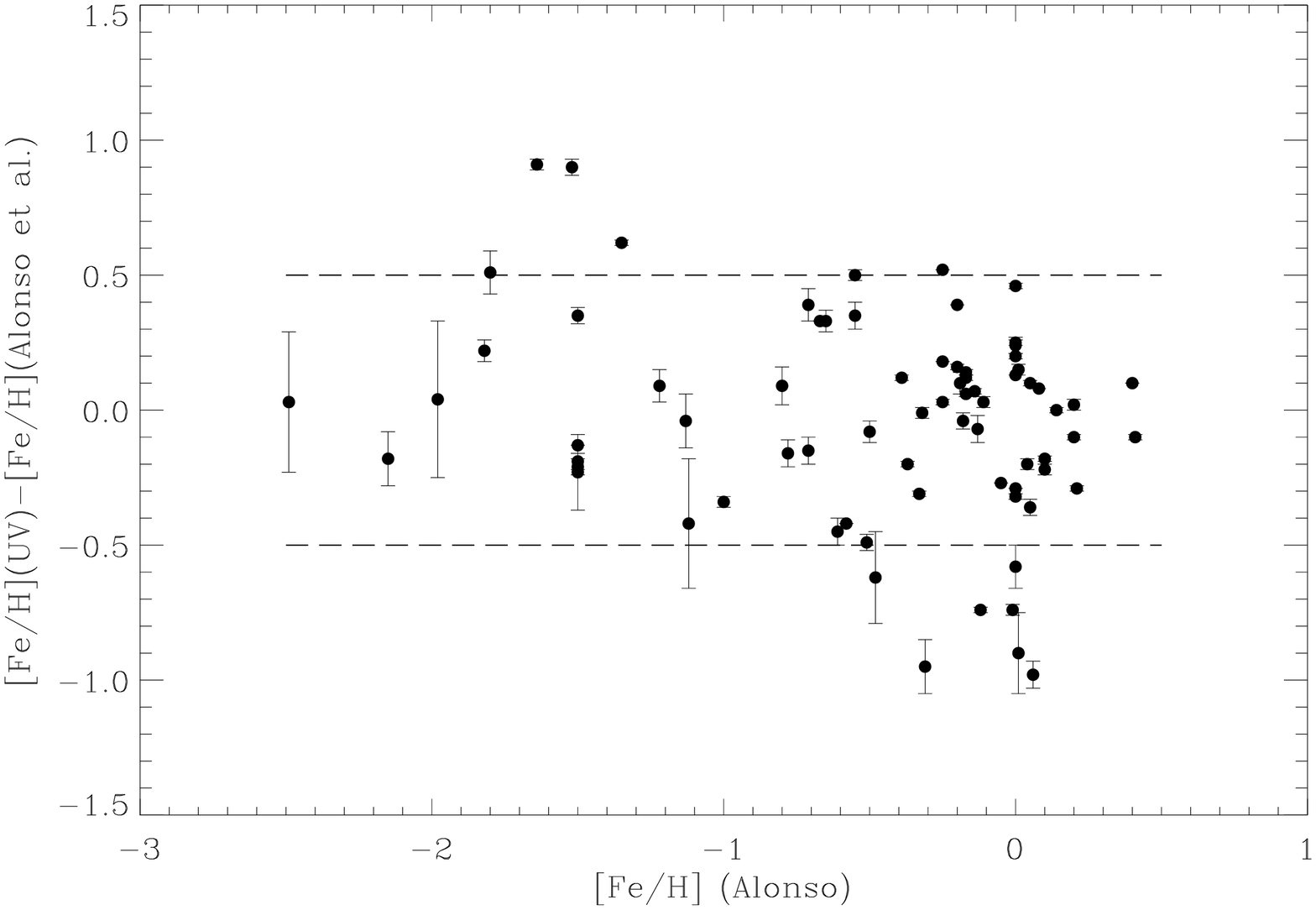}  
\protect\caption[]{{\it Upper panel}: relative differences between the
effective temperatures derived from the Infrared 
Flux Method (Alonso et al. 1996) and from
 the fit of the near-UV continuum (this work). The broken lines just
 indicate differences of  2\%. {\it Lower panel}: relative differences between
the metallicities compiled by Alonso et al. from the Cayrel et al. (1992) 
catalog and photometric calibrations with those derived from the near-UV
continuum. The broken lines just indicate differences of 0.5 dex.
\label{fig6}}
\end{figure}

Fig. 6 (middle panel) compares the metallicities listed by Alonso et
al. with those derived from the fit of the near-UV. Alonso et al. got
metallicity estimates from the catalogue gathered by Cayrel de Strobel
et al. (1992) for part of the sample, and completed the work using
photometric calibrations (Carney 1979, Schuster \& Nissen 1989). The
near-UV metallicities are on the same scale, as indicated by the mere
$-0.06$ dex mean difference, and the standard deviation is 0.4 dex,
which might well be entirely accounted for by the highly inhomogeneous
origin of the Alonso et al's  metallicities, i.e. our test may not reveal the true accuracy of the [Fe/H] estimates from the near-UV fluxes.

\subsection{Comparison with the stars analyzed by Gratton et al. (1996)}

Starting from  color-$T_{\rm eff}$ calibrations based on
published IRFM $T_{\rm eff}$s for solar-metallicity stars, Gratton et
al. (1996) derived consistent stellar parameters by requiring Kurucz's
model atmospheres to reproduce the iron ionization equilibrium.  They
noticed that  it was  not possible to completely zero the trends of the
iron abundance derived from lines
 with different excitation potentials, and keep the $T_{\rm eff}$s
consistent with the  IRFM photometric calibrations. 
Comparison of their ionization-equilibrium gravities with those
estimated by Allende Prieto et al. (1999) based on {\it Hipparcos}
parallaxes has shown  a  significant trend of the
difference with metallicity. However, such a trend
 is difficult to interpret, as many external elements, such as different 
$T_{\rm eff}$ scales, are at play  (see Allende Prieto et al.).

\begin{figure}[ht!]
\centering
\includegraphics[width=5.cm,angle=90]{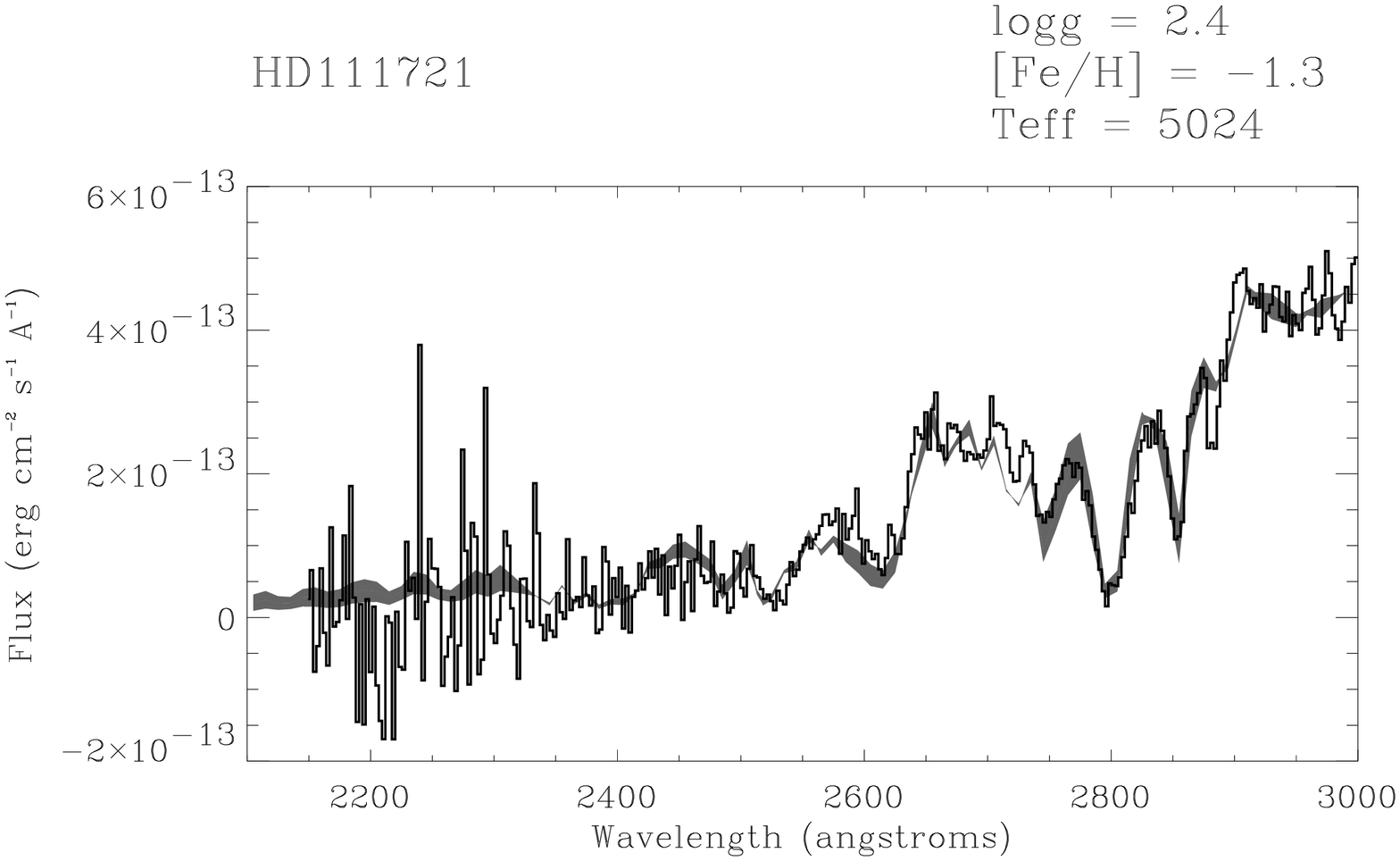}  
\includegraphics[width=5.cm,angle=90]{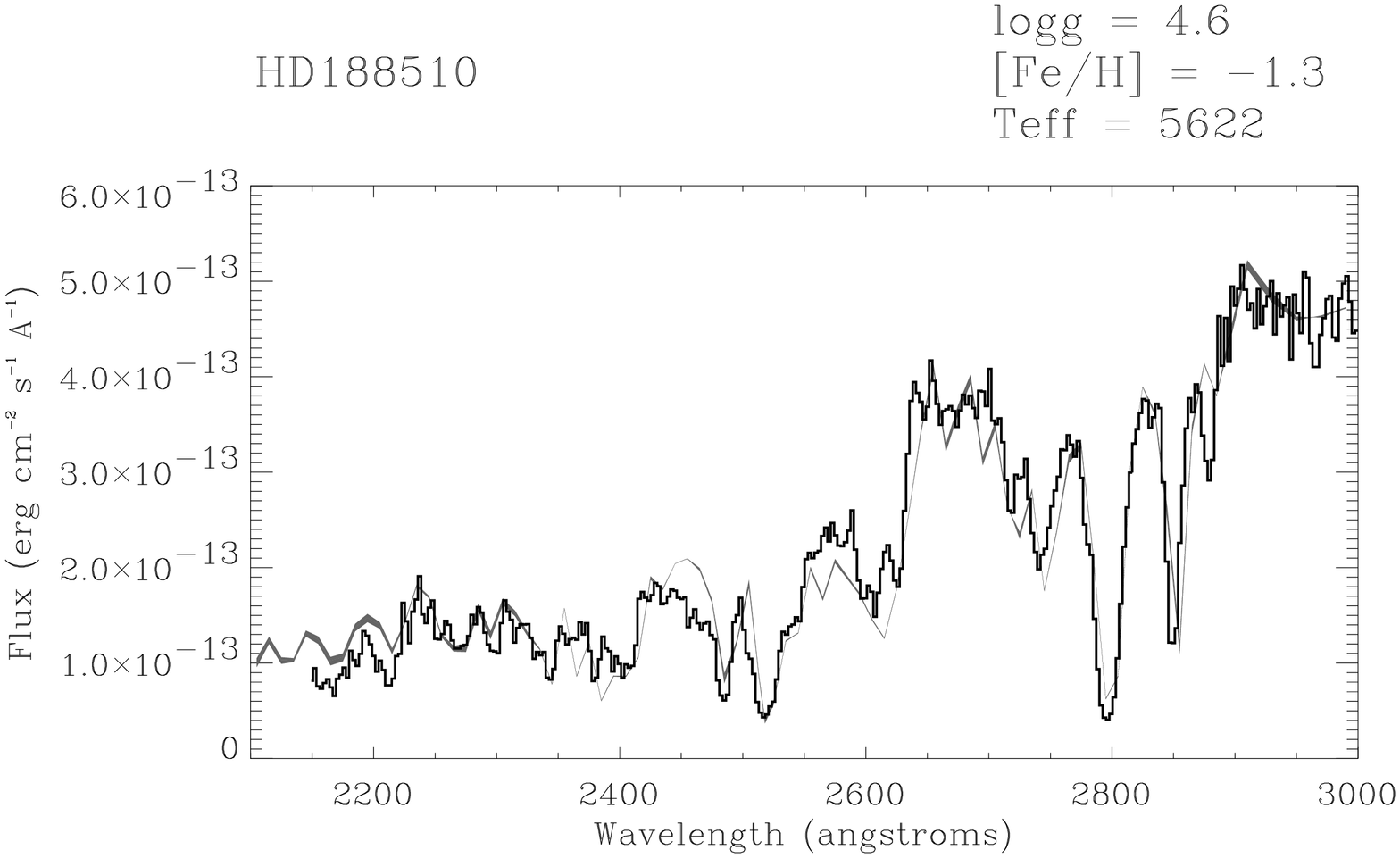}  
\includegraphics[width=5.cm,angle=90]{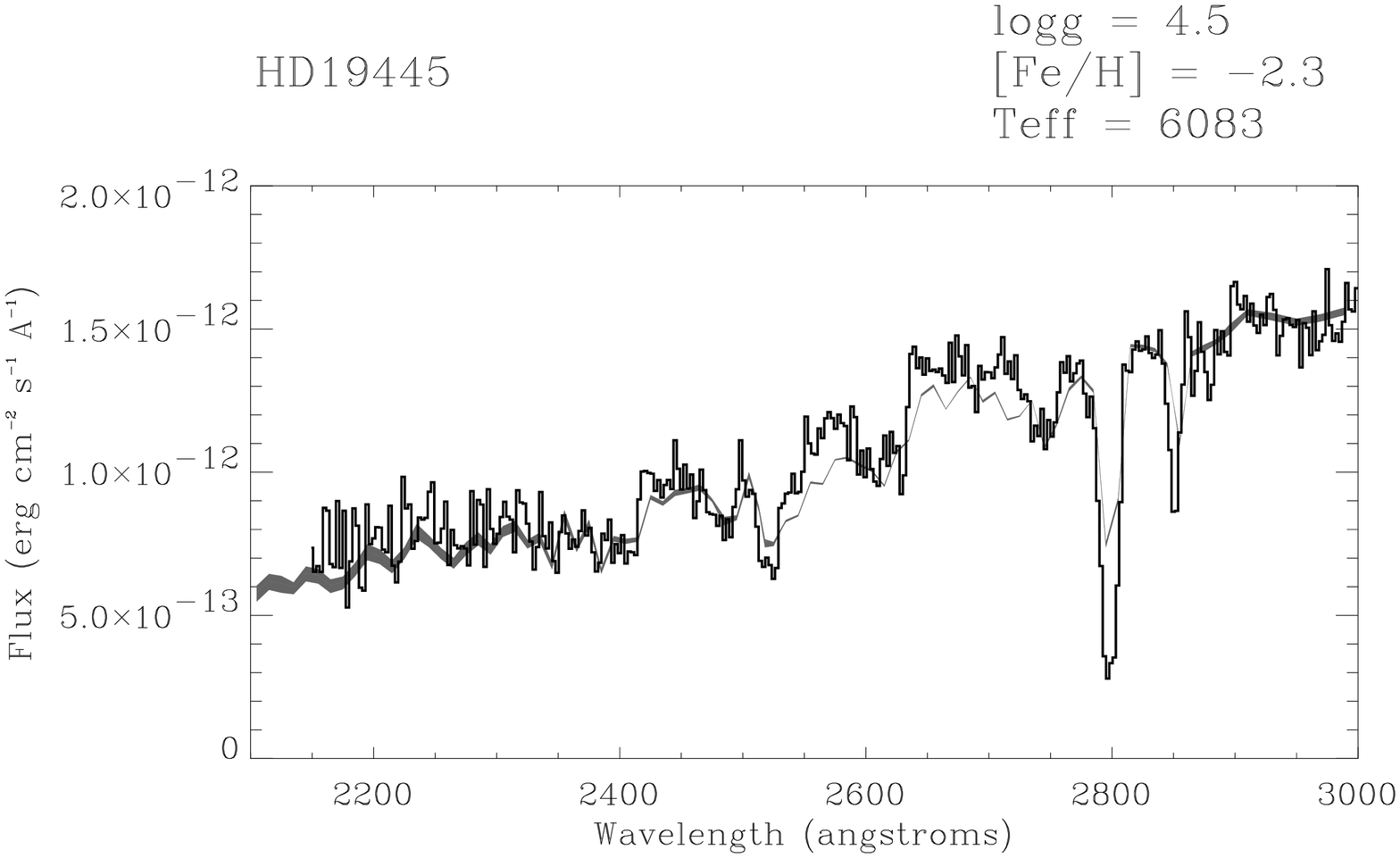}  
\includegraphics[width=5.cm,angle=90]{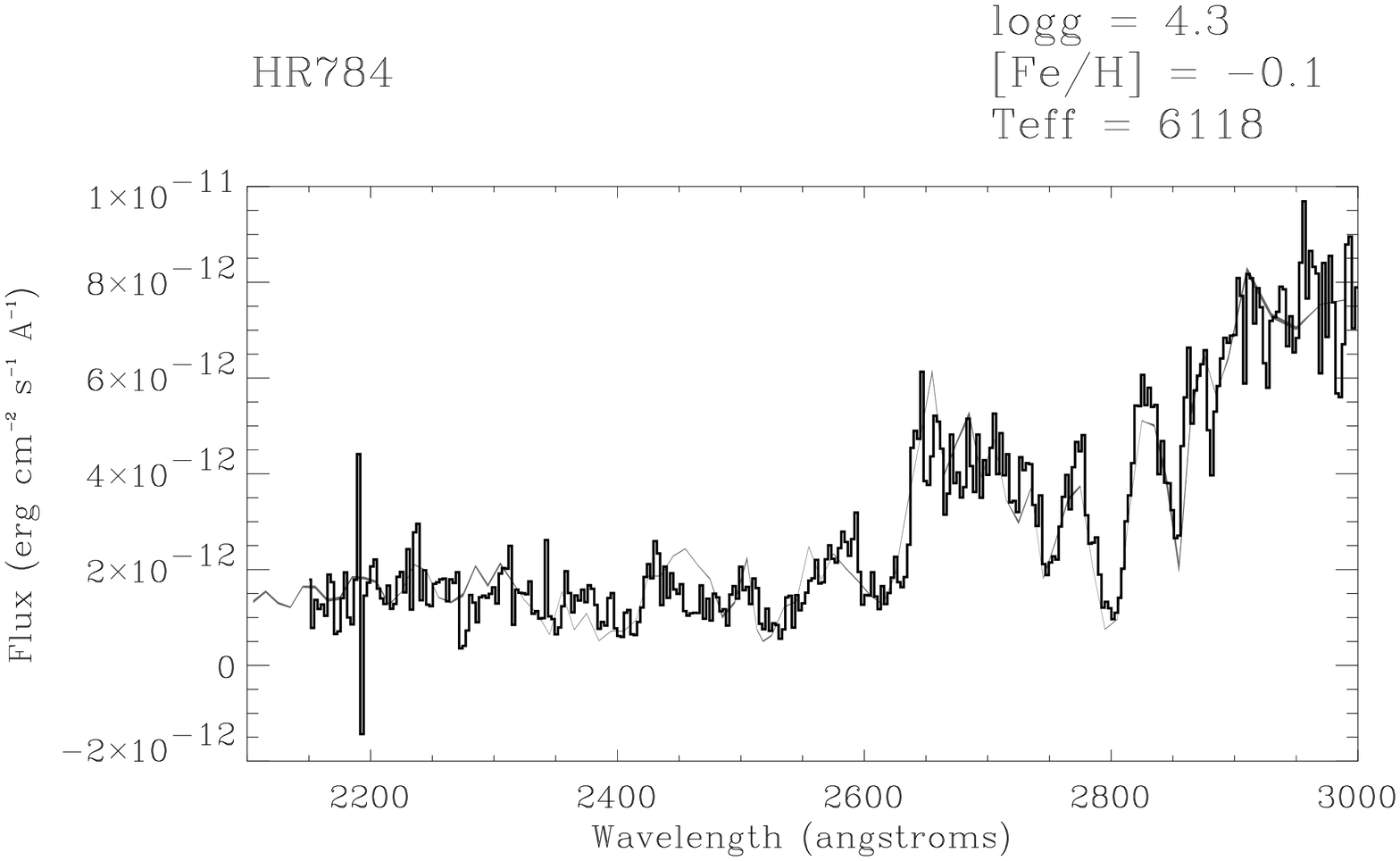}  
\includegraphics[width=5.cm,angle=90]{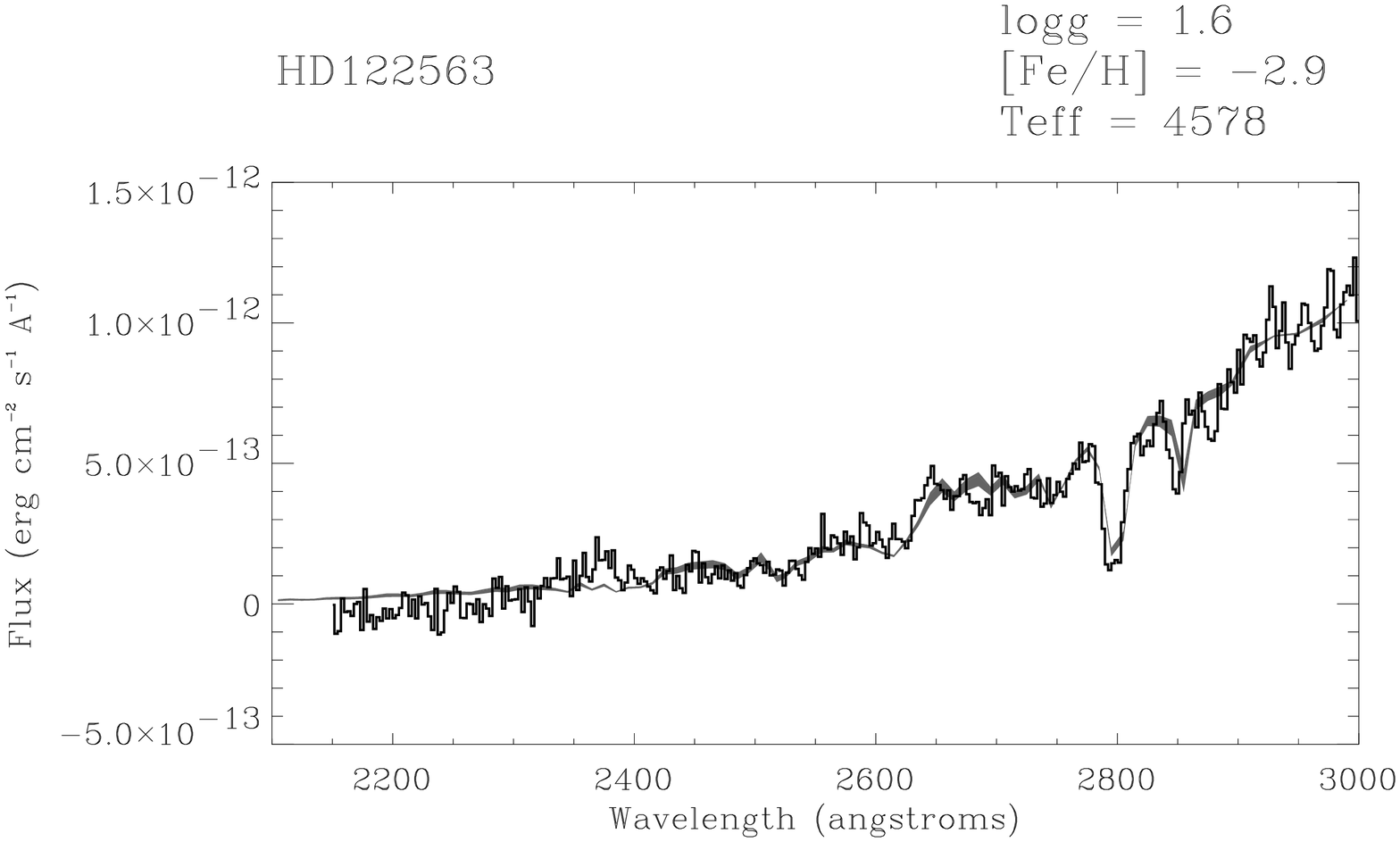}  
\includegraphics[width=5.cm,angle=90]{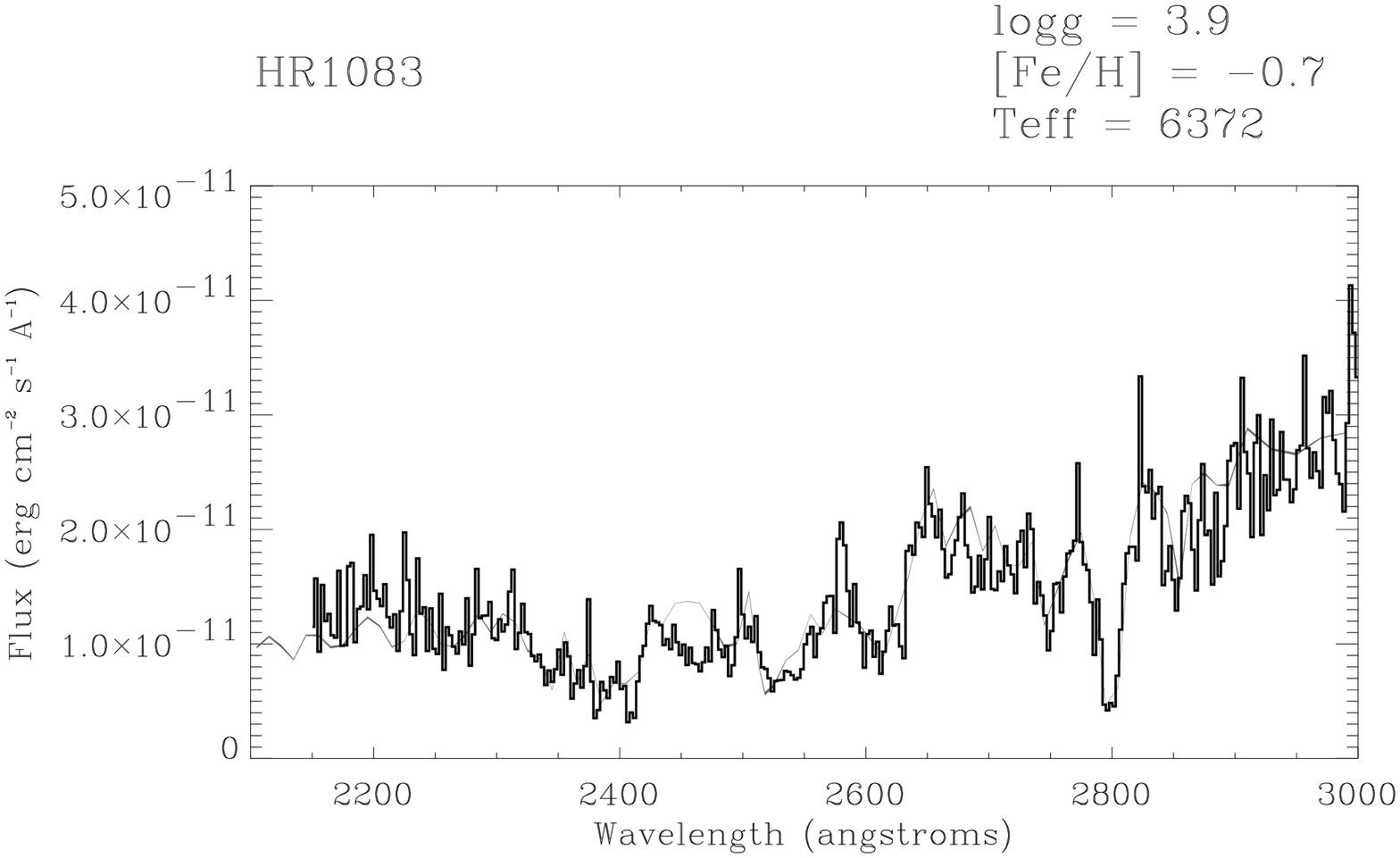}  
\protect\caption[]{Observed (thick solid line) and theoretical (shaded and
broken lines) fluxes at Earth for six of the stars analyzed by Gratton
et al. (1996).  The thickness of the  shaded lines represents the range
of possible fits resulting from uncertainties in  the derived gravity
and dilution factor $(pR)^2$.
\label{fig7}}
\end{figure}

Among several comparisons  performed by Gratton et al. to check their
 adopted photometric calibrations, they show the existence of a large
discrepancy between their $T_{\rm eff}$s and those derived by
Edvardsson et al. (1993) and Nissen et al. (1994), which strongly
correlates with the stellar metallicity. They find  that differences
between the atmospheric structures employed are the reason for the
discrepancy.  We are then interested in seeing whether Kurucz models,
and in particular, the calibrations based on IR fluxes of Kurucz models
obtained by Gratton et al.  are consistent with temperatures derived
from the near-UV fluxes. We found that 57 stars studied by Gratton et
al. had been observed by {\it Hipparcos} and the long-wavelength
cameras of the IUE at low dispersion.  The spectra of four of the stars
(HD108177, HD165195, HD187111, HD221170) could not be fitted by our
procedure.  Fig. 7 shows some comparisons between  observed (thick
solid lines) and synthetic spectra (shaded and broken lines). The
thickness of the shaded lines is used again to indicate the result of
using upper and lower limits of the flux dilution factor $(pR)^2$  in
the fit.

\begin{figure}[ht!]
\centering
\includegraphics[width=8.cm,angle=90]{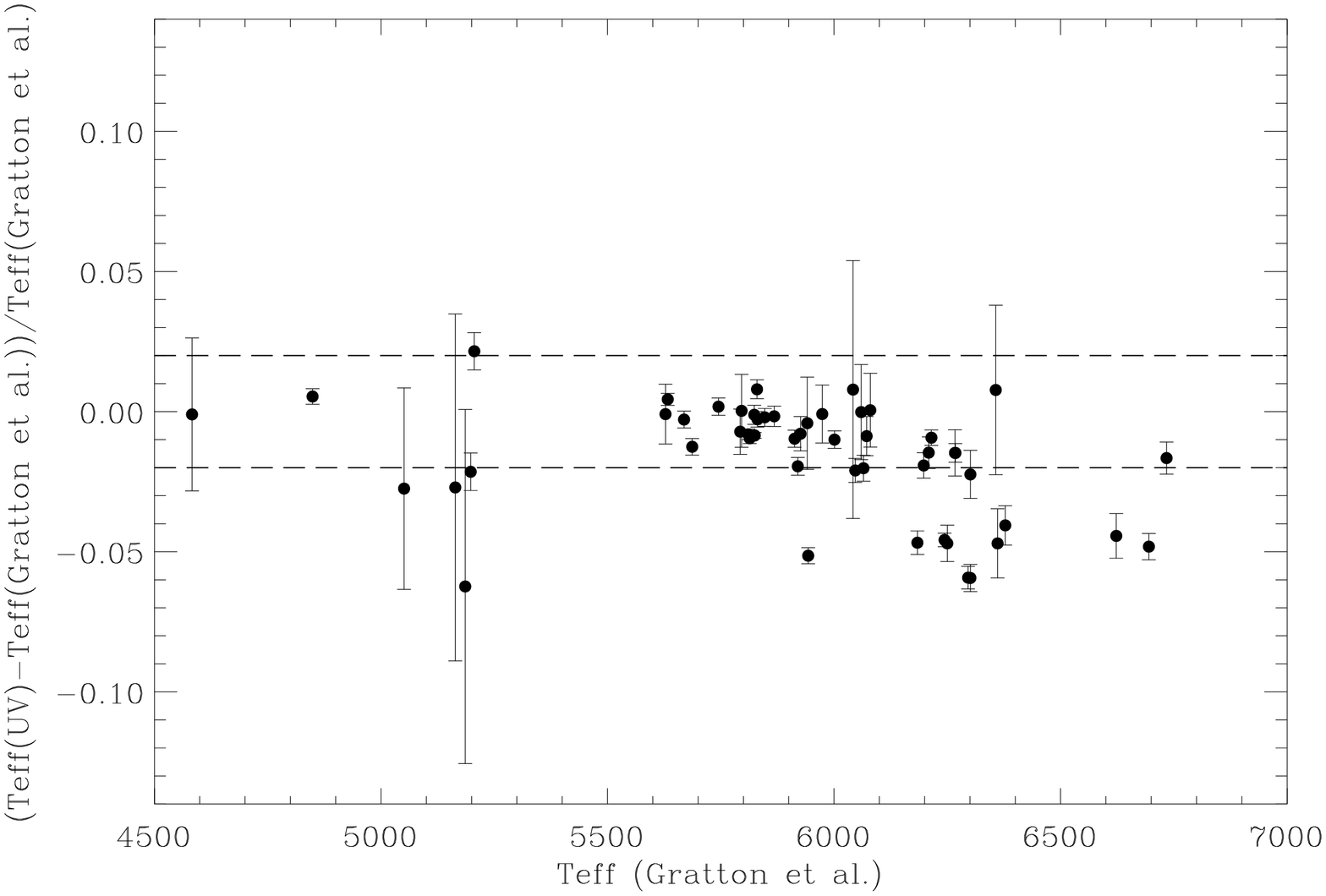}  
\includegraphics[width=8.cm,angle=90]{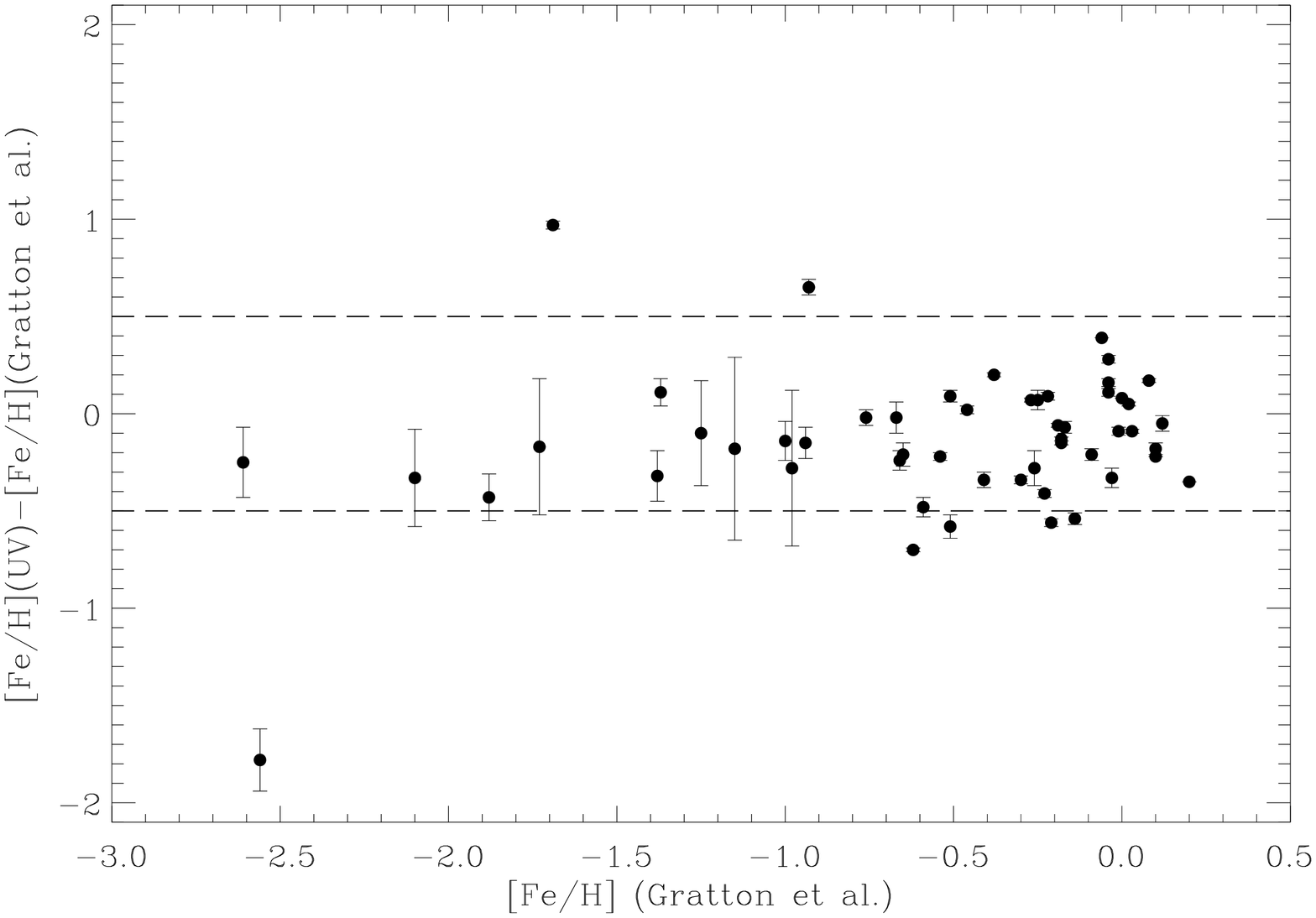}  
\protect\caption[]{{\it Upper panel}: relative differences between the
effective temperatures derived by  Gratton et al. (1996) and those obtained
in this work from 
 the fit of the near-UV continuum. The broken lines just
 indicate differences of  2\%. {\it Lower panel}: relative differences between
the metallicities derived by Gratton  et al.  with those derived from 
the near-UV continuum. The broken lines just indicate 
differences of 0.5 dex.
\label{fig8}}
\end{figure}

 Fig. 8 (upper panel) shows the comparison between the retrieved
$T_{\rm eff}$s and those published by Gratton et al. (1996). The
standard deviation of the two $T_{\rm eff}$ scales is a mere 2\%,
although it is apparent, as  was found for the comparison with Alonso
et al., that the $T_{\rm eff}$s from the near-UV fluxes for stars
hotter than $\simeq 6200$ K are systematically smaller.  Restricting
the comparison to stars cooler than 6200 K the standard deviation is
reduced to 1.6\%. Large uncertainties affect the translation between
fluxes at the stellar surface and at Earth for the cooler stars,
 as indicated by the large error bars. Figure 8 (lower panel) shows
 agreement for the metallicity scales. Excluding  a particularly
 deviant case, HD184711, the mean difference is $-0.110 \pm 0.006$ and
 the standard deviation is 0.3 dex. No correlation is apparent between
the discrepancies in $T_{\rm eff}$ and  metallicity.

\section{Summary and conclusions}

The parallaxes measured by the {\it Hipparcos} mission provide a way
to translate the spectral energy distributions observed 
at Earth to absolute fluxes escaping
from the stellar surface. Opacities and models employed to compute the
predicted flux can therefore be  checked using not only the shape of
the continuum, but also its absolute value. 

Effective temperatures  derived
 by Alonso et al. (1996) using the Infrared Flux Method (Blackwell et
al. 1991)  are compared with those derived here from  absolute near-UV
fluxes observed by the IUE satellite. The study shows
 that for stars with $T_{\rm eff}$ in the range $4000-6000$ K, the two
methods provide concordant results.
 For stars cooler than 4000 K, Alonso et al.  have shown that the
Infrared Flux Method is especially sensitive to errors in the observed
quantities, and that might be the reason for the discrepancy with the
near-UV $T_{\rm eff}$s. The systematic differences found for stars
hotter than 6000 K may  reflect problems in the model atmospheres
and/or the opacities for those temperatures, although other effects can
not be ruled out at this stage.  The metallicities compiled by Alonso
et al. from  the Cayrel et al. (1992) catalogue and photometric calibrations
are in agreement with those retrieved from the analysis of  near-UV
spectra, at least within their expected uncertainties.  A similar
comparison is performed with the multi-criteria atmospheric parameters
derived by Gratton et al. (1996), strengthening the results  just
described.

Previous comparisons between synthetic and observed near-UV spectra for
late-G and early-K stars were performed by 
Morossi et al.  (1993; see also Malagnini et al. 1992). 
They used older IUE data but the same (or very similar) Kurucz models. 
Their approach was different, in the sense they used
 atmospheric parameters predetermined from the literature (spectroscopic
analysis) and empirical photometric 
calibrations to select a model and then compare
it with the observations.  In contrast to our
conclusions, they found strong discrepancies between observed and
predicted near-UV fluxes for several stars: 
predicted fluxes were smaller than observations.  
Whether systematic errors in the
stellar parameters or deficiencies in the older IUE fluxes were responsible 
for the failure is unclear. 

We conclude that Kurucz flux-constant model atmospheres
 are able to  reproduce the near-UV absolute continuum for stars with
$4000 \le T_{\rm eff}  \le 6000$ K. This holds for any metallicity
and gravity, although it is clearly worthwhile to concentrate future
efforts on the detailed study of obvious small discrepancies for
particular cases and particular wavelengths, as they should shed light
on important issues, such as chemical abundances of several elements
which produce features in the considered spectral range (e.g. boron;
Cunha \& Smith 1999).  The retrieved $T_{\rm eff}$s and [Fe/H]s are in
excellent agreement with other reliable spectroscopic and photometric
indicators, which we interpret as an important success of the models
indicating that:  i) the average temperature stratification  in the
layers $0 \le \log \tau \le 1$ is appropriate, ii) the fundamental
hypotheses employed to construct the models are adequate to interpret
the near-UV continuum, and iii) the line and continuum opacities in the
UV are essentially understood.  The newer version of the IUE final
archive (INES) and the application of  recently-suggested  procedures
  (Massa \& Fitzpatrick 1998) in order to improve the quality of IUE
fluxes will provide an excellent opportunity to check and extend the
analyses presented here, as well as to exploit  the wealth of
 information  coded in the near-UV continuum.

\acknowledgments

We are indebted to the referee, Derck Massa, for many interesting
comments that helped to improve the paper.   Ivan Hubeny is thanked for
estimulating discussions. This work has been partially funded by the NSF
(grant AST961814) and the Robert A. Welch Foundation of Houston, Texas.
We have made use of data from the IUE Final Archive at VILSPA,  the
{\it Hipparcos} astrometric mission of the ESA,  the NASA ADS, and the
CDS service  for astronomical catalogues.

\clearpage

%% No more than seven \figcaption commands are allowed per page,
%% so if you have more than seven captions, insert a \clearpage

%% There must be a \figcaption command for each legend. Key the text of the
%% legend and the optional \label in curly braces. If you wish, you may
%% include the name of the corresponding figure file in square brackets.
%% The label is for identification purposes only. It will not insert the
%% figures themselves into the document.
%% If you want to include your art in the paper, use \plotone.
%% Refer to the on-line documentation for details.

\clearpage

%table 1

\begin{deluxetable}{lllllll}
\small
\tablecaption{Data for the stars in the comparison with Alonso et al. (1996)
\label{table1}}
\tablehead{
\colhead{Star}  & \colhead{$T_{\rm eff}^{UV}$}& \colhead{$T_{\rm eff}^{IRFM}$} & \colhead{[Fe/H]$^{UV}$} & 
\colhead{[Fe/H]$^{Lit}$} &  \colhead{$p$} &
\colhead{$\log g$}  \\
 &  \colhead{K} & \colhead{K} & \colhead{dex} & \colhead{dex} & \colhead{mas}  & \colhead{dex} }
\startdata 
 G013-035 &   6146$^{+222}_{-230}$  & 6097 & $-1.94^{+0.29}_{-0.31}$ & $-1.98$ &     10.95$\pm   1.29$ & 4.45$\pm 0.12$  \\
 G014-006 &   4012$^{+18}_{+1}$  & 3748 & $-0.12^{+0.02}_{+0.05}$ & $+0.10$ &     92.75$\pm   0.96$ & 4.70$\pm 0.07$  \\
 G019-013 &   4325$^{+4}_{-1}$  & 4314 & $-0.29^{+0.00}_{+0.02}$ & $+0.00$ &     92.98$\pm   1.04$ & 4.66$\pm 0.07$  \\
 G019-024 &   4368$^{+7}_{-8}$  & 4065 & $+0.50^{+0.00}_{-0.00}$ & $+0.40$ &    129.54$\pm   0.95$ & 4.80$\pm 0.07$  \\
 G025-015 &   5650$^{+125}_{-106}$  & 5747 & $-1.10^{+0.17}_{-0.14}$ & $-0.48$ &     17.83$\pm   1.29$ & 3.81$\pm 0.09$  \\
 G043-003 &   6389$^{+189}_{-195}$  & 6330 & $-2.46^{+0.26}_{-0.30}$ & $-2.49$ &     12.44$\pm   1.06$ & 4.14$\pm 0.10$  \\
 G058-025 &   6028$^{+100}_{-119}$  & 6001 & $-1.73^{+0.14}_{-0.17}$ & $-1.50$ &     19.23$\pm   1.13$ & 4.35$\pm 0.09$  \\
 G063-009 &   5886$^{+94}_{-96}$  & 5884 & $-0.71^{+0.07}_{-0.11}$ & $-0.80$ &     24.65$\pm   1.44$ & 4.19$\pm 0.09$  \\
 G080-015 &   5810$^{+34}_{-34}$  & 5798 & $-0.86^{+0.05}_{-0.05}$ & $-0.71$ &     41.07$\pm   0.86$ & 4.35$\pm 0.07$  \\
 G090-025 &   5456$^{+36}_{-36}$  & 5441 & $-1.60^{+0.04}_{-0.05}$ & $-1.82$ &     35.29$\pm   1.04$ & 4.63$\pm 0.07$  \\
 G112-054 &   5282$^{+30}_{-45}$  & 5134 & $-0.32^{+0.04}_{-0.10}$ & $-0.65$ &     52.01$\pm   1.85$ & 4.60$\pm 0.08$  \\
 G182-007 &   5303$^{+49}_{-49}$  & 5175 & $-0.09^{+0.04}_{-0.05}$ & $-0.19$ &     19.78$\pm   1.07$ & 4.16$\pm 0.08$  \\
 G182-019 &   5810$^{+64}_{-71}$  & 5771 & $-0.32^{+0.06}_{-0.10}$ & $-0.71$ &     18.32$\pm   0.78$ & 4.33$\pm 0.08$  \\
 G184-029 &   3935$^{+16}_{-17}$  & 3760 & $-1.15^{-0.03}_{+0.02}$ & $-1.50$ &     58.60$\pm   1.60$ & 4.58$\pm 0.07$  \\
 G191-051 &   4023$^{+27}_{-25}$  & 3832 & $+0.46^{+0.01}_{-0.05}$ & $+0.00$ &     80.13$\pm   1.67$ & 4.88$\pm 0.07$  \\
 G196-009 &   3860$^{+2}_{-0}$  & 3764 & $-1.34^{-0.02}_{+0.03}$ & $-1.00$ &    205.22$\pm   0.81$ & 4.61$\pm 0.07$  \\
 G200-062 &   5150$^{+12}_{-11}$  & 5098 & $-0.05^{+0.02}_{-0.01}$ & $-0.55$ &     41.83$\pm   0.63$ & 4.52$\pm 0.07$  \\
 G237-062 &   5265$^{+19}_{-50}$  & 5323 & $-0.27^{+0.01}_{-0.09}$ & $-0.39$ &     23.16$\pm   0.67$ & 4.23$\pm 0.07$  \\
 G244-059 &   5523$^{+45}_{-59}$  & 5501 & $-0.20^{+0.05}_{-0.07}$ & $-0.55$ &     25.82$\pm   1.07$ & 4.48$\pm 0.08$  \\
 GJ782 &   3966$^{+27}_{-18}$  & 3900 & $-1.71^{+0.03}_{+0.02}$ & $-1.50$ &     63.82$\pm   1.49$ & 4.61$\pm 0.07$  \\
 GJ820B &   3845$^{+4}_{-7}$  & 3786 & $-1.63^{+0.00}_{-0.01}$ & $-1.50$ &    285.42$\pm   0.72$ & 4.65$\pm 0.07$  \\
 GJ884 &   3834$^{+10}_{-11}$  & 3746 & $-1.69^{+0.03}_{-0.02}$ & $-1.50$ &    122.80$\pm   0.94$ & 4.64$\pm 0.07$  \\
 HD103095 &   5069$^{+4}_{-2}$  & 5029 & $-0.73^{+0.01}_{-0.00}$ & $-1.35$ &    109.21$\pm   0.78$ & 4.68$\pm 0.07$  \\
 HD111980 &   5697$^{+173}_{-161}$  & 5624 & $-1.54^{+0.24}_{-0.17}$ & $-1.12$ &     12.48$\pm   1.38$ & 3.97$\pm 0.12$  \\
 HD118100 &   4127$^{+15}_{-13}$  & 4179 & $-2.00^{+0.01}_{+0.01}$ & $-0.07$ &     50.54$\pm   0.99$ & 4.62$\pm 0.07$  \\
 HD134439 &   5110$^{+31}_{-32}$  & 4974 & $-0.62^{+0.03}_{-0.03}$ & $-1.52$ &     34.14$\pm   1.36$ & 4.74$\pm 0.08$  \\
 HD157089 &   5532$^{+34}_{-32}$  & 5662 & $-1.00^{+0.00}_{-0.01}$ & $-0.58$ &     25.88$\pm   0.95$ & 4.01$\pm 0.08$  \\
 HD188510 &   5597$^{+58}_{-57}$  & 5564 & $-1.29^{+0.08}_{-0.08}$ & $-1.80$ &     25.32$\pm   1.17$ & 4.63$\pm 0.08$  \\
 HD193901 &   5777$^{+73}_{-84}$  & 5750 & $-1.17^{+0.10}_{-0.11}$ & $-1.13$ &     22.88$\pm   1.24$ & 4.57$\pm 0.08$  \\
 HD19445 &   6065$^{+75}_{-82}$  & 6050 & $-2.33^{+0.10}_{-0.15}$ & $-2.15$ &     25.85$\pm   1.14$ & 4.51$\pm 0.08$  \\
 HD201891 &   5929$^{+64}_{-62}$  & 5909 & $-1.13^{+0.06}_{-0.07}$ & $-1.22$ &     28.26$\pm   1.01$ & 4.33$\pm 0.08$  \\
 HD25329 &   4870$^{+14}_{-12}$  & 4842 & $-0.73^{+0.02}_{-0.01}$ & $-1.64$ &     54.14$\pm   1.08$ & 4.78$\pm 0.07$  \\
 HD4307 &   5726$^{+51}_{-53}$  & 5753 & $-0.20^{+0.05}_{-0.03}$ & $-0.13$ &     31.39$\pm   1.03$ & 3.97$\pm 0.08$  \\
 HR0660 &   5611$^{+12}_{-12}$  & 5591 & $-0.64^{+0.01}_{-0.01}$ & $-0.33$ &     92.20$\pm   0.84$ & 4.30$\pm 0.07$  \\
 HR1325 &   5080$^{+2}_{-0}$  & 5040 & $-0.05^{+0.01}_{+0.00}$ & $-0.17$ &    198.24$\pm   0.84$ & 4.51$\pm 0.07$  \\
 HR1543 &   6331$^{+22}_{-21}$  & 6482 & $-0.16^{+0.02}_{-0.02}$ & $+0.04$ &    124.60$\pm   0.95$ & 4.16$\pm 0.07$  \\
 HR1729 &   5826$^{+22}_{-16}$  & 5847 & $+0.24^{+0.03}_{-0.01}$ & $+0.00$ &     79.08$\pm   0.90$ & 4.19$\pm 0.07$  \\
 HR1925 &   5179$^{+7}_{-13}$  & 5185 & $+0.19^{+0.00}_{-0.03}$ & $-0.20$ &     81.69$\pm   0.83$ & 4.51$\pm 0.07$  \\
 HR2085 &   6524$^{+42}_{-51}$  & 7013 & $-0.92^{+0.05}_{-0.08}$ & $+0.06$ &     66.47$\pm   0.74$ & 3.79$\pm 0.07$  \\
 HR219 &   5809$^{+7}_{-6}$  & 5817 & $-0.32^{+0.01}_{-0.01}$ & $+0.00$ &    167.99$\pm   0.62$ & 4.33$\pm 0.07$  \\
 HR2852 &   6470$^{+67}_{-52}$  & 7020 & $-1.26^{+0.10}_{-0.05}$ & $-0.31$ &     54.06$\pm   0.95$ & 3.78$\pm 0.07$  \\
 HR321 &   5386$^{+0}_{-3}$  & 5315 & $-0.34^{+0.00}_{-0.00}$ & $-0.67$ &    132.40$\pm   0.60$ & 4.56$\pm 0.07$  \\
 HR3262 &   6115$^{+36}_{-34}$  & 6242 & $-0.58^{+0.04}_{-0.03}$ & $-0.50$ &     55.17$\pm   0.93$ & 4.11$\pm 0.07$  \\
 HR3775 &   5946$^{+25}_{-22}$  & 6338 & $-0.75^{+0.02}_{-0.01}$ & $-0.01$ &     74.15$\pm   0.74$ & 3.51$\pm 0.07$  \\
 HR4421 &   6356$^{+51}_{-50}$  & 6634 & $-1.06^{+0.05}_{-0.05}$ & $-0.61$ &     30.40$\pm   0.60$ & 3.87$\pm 0.07$  \\
 HR4496 &   5371$^{+9}_{-6}$  & 5342 & $-0.07^{+0.01}_{-0.00}$ & $-0.14$ &    104.81$\pm   0.72$ & 4.46$\pm 0.07$  \\
 HR4540 &   5955$^{+19}_{-16}$  & 6095 & $-0.08^{+0.01}_{-0.00}$ & $+0.21$ &     91.74$\pm   0.77$ & 3.95$\pm 0.07$  \\
 HR4657 &   6136$^{+51}_{-57}$  & 6208 & $-0.94^{+0.05}_{-0.10}$ & $-0.78$ &     44.34$\pm   1.01$ & 4.30$\pm 0.07$  \\
 HR4785 &   5784$^{+10}_{-11}$  & 5867 & $-0.22^{+0.01}_{-0.02}$ & $-0.25$ &    119.46$\pm   0.83$ & 4.34$\pm 0.07$  \\
 HR483 &   5755$^{+15}_{-21}$  & 5874 & $-0.08^{+0.02}_{-0.02}$ & $-0.11$ &     79.09$\pm   0.83$ & 4.26$\pm 0.07$  \\
 HR493 &   5072$^{+4}_{-3}$  & 5172 & $-0.04^{+0.01}_{-0.01}$ & $-0.20$ &    133.91$\pm   0.91$ & 4.51$\pm 0.07$  \\
 HR4983 &   5857$^{+12}_{-11}$  & 5964 & $-0.08^{+0.01}_{-0.01}$ & $+0.10$ &    109.23$\pm   0.72$ & 4.29$\pm 0.07$  \\
 HR5447 &   6405$^{+36}_{-25}$  & 6707 & $-1.00^{+0.03}_{-0.01}$ & $-0.51$ &     64.66$\pm   0.72$ & 4.01$\pm 0.07$  \\
 HR5534 &   5946$^{+24}_{-24}$  & 6019 & $+0.10^{+0.01}_{-0.01}$ & $+0.20$ &     55.73$\pm   0.80$ & 4.37$\pm 0.07$  \\
 HR5568 &   4632$^{+3}_{-2}$  & 4605 & $+0.16^{-0.02}_{+0.00}$ & $+0.01$ &    169.32$\pm   1.67$ & 4.61$\pm 0.07$  \\
 HR5634 &   6406$^{+42}_{-41}$  & 6571 & $-0.31^{+0.03}_{-0.04}$ & $+0.05$ &     50.70$\pm   0.76$ & 4.09$\pm 0.07$  \\
 HR5758 &   6434$^{+127}_{-122}$  & 6831 & $-0.89^{+0.15}_{-0.12}$ & $+0.01$ &     19.73$\pm   0.92$ & 3.86$\pm 0.08$  \\
 HR5868 &   5861$^{+15}_{-15}$  & 5897 & $+0.15^{+0.01}_{-0.01}$ & $+0.05$ &     85.08$\pm   0.80$ & 4.17$\pm 0.07$  \\
 HR5901 &   4798$^{+7}_{-6}$  & 4811 & $+0.20^{-0.01}_{+0.01}$ & $+0.00$ &     32.13$\pm   0.61$ & 3.06$\pm 0.07$  \\
 HR5914 &   5763$^{+13}_{-14}$  & 5774 & $-0.57^{+0.01}_{-0.02}$ & $-0.37$ &     63.08$\pm   0.54$ & 3.97$\pm 0.07$  \\
 HR5933 &   6154$^{+20}_{-20}$  & 6233 & $-0.33^{+0.02}_{-0.02}$ & $-0.32$ &     89.92$\pm   0.72$ & 4.09$\pm 0.07$  \\
 HR5968 &   5770$^{+16}_{-18}$  & 5777 & $-0.11^{+0.01}_{-0.02}$ & $-0.17$ &     57.38$\pm   0.71$ & 4.18$\pm 0.07$  \\
 HR6556 &   7325$^{+28}_{-31}$  & 7923 & $-0.58^{-0.08}_{+0.07}$ & $+0.00$ &     69.84$\pm   0.88$ & 3.35$\pm 0.07$  \\
 HR6752 &   5023$^{+0}_{-0}$  & 4978 & $-0.03^{-0.01}_{+0.01}$ & $-0.17$ &    196.62$\pm   1.38$ & 4.37$\pm 0.07$  \\
 HR6806 &   5034$^{+0}_{-1}$  & 4947 & $+0.27^{+0.00}_{+0.00}$ & $-0.25$ &     90.11$\pm   0.54$ & 4.57$\pm 0.07$  \\
 HR72 &   5628$^{+26}_{-26}$  & 5683 & $+0.22^{+0.02}_{-0.02}$ & $+0.20$ &     42.67$\pm   0.85$ & 4.28$\pm 0.07$  \\
 HR7373 &   5436$^{+14}_{-17}$  & 5518 & $+0.31^{+0.01}_{-0.02}$ & $+0.41$ &     66.01$\pm   0.77$ & 4.10$\pm 0.07$  \\
 HR7462 &   5143$^{+3}_{-0}$  & 5227 & $-0.07^{+0.00}_{+0.00}$ & $-0.25$ &    173.41$\pm   0.46$ & 4.52$\pm 0.07$  \\
 HR7503 &   5605$^{+13}_{-12}$  & 5763 & $+0.14^{+0.01}_{+0.00}$ & $+0.14$ &     46.25$\pm   0.50$ & 4.17$\pm 0.07$  \\
 HR7504 &   5676$^{+16}_{-17}$  & 5767 & $+0.16^{+0.00}_{-0.02}$ & $+0.08$ &     46.70$\pm   0.52$ & 4.29$\pm 0.07$  \\
 HR7914 &   5795$^{+17}_{-16}$  & 5761 & $+0.13^{+0.00}_{+0.00}$ & $+0.00$ &     47.65$\pm   0.76$ & 4.41$\pm 0.07$  \\
 HR8085 &   4402$^{+3}_{+2}$  & 4323 & $-0.32^{+0.00}_{+0.01}$ & $-0.05$ &    287.13$\pm   1.51$ & 4.67$\pm 0.07$  \\
 HR8086 &   4092$^{-1}_{+1}$  & 3865 & $-0.22^{-0.03}_{+0.03}$ & $-0.18$ &    285.42$\pm   0.72$ & 4.73$\pm 0.07$  \\
 HR8832 &   4809$^{+0}_{-0}$  & 4785 & $+0.25^{+0.01}_{+0.01}$ & $+0.00$ &    153.24$\pm   0.65$ & 4.58$\pm 0.07$  \\
 HR8905 &   5441$^{+52}_{-53}$  & 5954 & $-0.86^{+0.01}_{-0.01}$ & $-0.12$ &     18.83$\pm   0.72$ & 2.61$\pm 0.08$  \\
\enddata
\end{deluxetable}

%table 2

\begin{deluxetable}{lllllll}
\small
\tablecaption{Data for the stars in the comparison with Gratton et al. (1996)
\label{table2}}
\tablehead{
\colhead{Star}  & \colhead{$T_{\rm eff}^{UV}$}& \colhead{$T_{\rm eff}^{Spec}$} & \colhead{[Fe/H]$^{UV}$} & 
\colhead{[Fe/H]$^{Spec}$} &  \colhead{$p$} &
\colhead{$\log g$}  \\
 &  \colhead{K} & \colhead{K} & \colhead{dex} & \colhead{dex} & \colhead{mas}  & \colhead{dex} }
\startdata 
 HD111721 &   5024$^{+319}_{-486}$  & 5164 & $-1.26^{+0.40}_{-0.99}$ & $-0.98$ &      3.29$\pm   1.11$ & 2.35$\pm 0.30$  \\
 HD114762 &   5916$^{+97}_{-100}$  & 5941 & $-0.69^{+0.08}_{-0.10}$ & $-0.67$ &     24.65$\pm   1.44$ & 4.21$\pm 0.09$  \\
 HD114946 &   5086$^{+34}_{-60}$  & 5198 & $+0.07^{+0.04}_{-0.16}$ & $+0.12$ &     25.89$\pm   0.73$ & 3.18$\pm 0.07$  \\
 HD122563 &   4578$^{+125}_{-144}$  & 4583 & $-2.86^{+0.18}_{-0.29}$ & $-2.61$ &      3.76$\pm   0.72$ & 1.58$\pm 0.18$  \\
 HD160617 &   6089$^{+277}_{-293}$  & 6042 & $-1.90^{+0.35}_{-0.36}$ & $-1.73$ &      8.66$\pm   1.25$ & 3.90$\pm 0.14$  \\
 HD166161 &   4862$^{+327}_{-506}$  & 5186 & $-1.33^{+0.47}_{-1.50}$ & $-1.15$ &      3.25$\pm   1.19$ & 2.35$\pm 0.33$  \\
 HD184711 &   4064$^{+172}_{-287}$  & 4157 & $-4.34^{+0.16}_{-0.20}$ & $-2.56$ &      3.15$\pm   1.16$ & 1.97$\pm 0.33$  \\
 HD188510 &   5622$^{+59}_{-58}$  & 5628 & $-1.26^{+0.07}_{-0.08}$ & $-1.37$ &     25.32$\pm   1.17$ & 4.64$\pm 0.08$  \\
 HD193901 &   5797$^{+75}_{-83}$  & 5796 & $-1.14^{+0.10}_{-0.12}$ & $-1.00$ &     22.88$\pm   1.24$ & 4.58$\pm 0.08$  \\
 HD19445 &   6083$^{+80}_{-73}$  & 6080 & $-2.31^{+0.12}_{-0.08}$ & $-1.88$ &     25.85$\pm   1.14$ & 4.52$\pm 0.08$  \\
 HD201891 &   5968$^{+61}_{-67}$  & 5974 & $-1.09^{+0.08}_{-0.07}$ & $-0.94$ &     28.26$\pm   1.01$ & 4.35$\pm 0.08$  \\
 HD208906 &   6018$^{+42}_{-42}$  & 6072 & $-0.86^{+0.06}_{-0.05}$ & $-0.65$ &     34.12$\pm   0.70$ & 4.36$\pm 0.07$  \\
 HD22879 &   5879$^{+36}_{-35}$  & 5926 & $-0.78^{+0.04}_{-0.03}$ & $-0.76$ &     41.07$\pm   0.86$ & 4.37$\pm 0.07$  \\
 HD25329 &   4875$^{+13}_{-13}$  & 4849 & $-0.72^{+0.02}_{-0.02}$ & $-1.69$ &     54.14$\pm   1.08$ & 4.78$\pm 0.07$  \\
 HD44007 &   4912$^{+181}_{-240}$  & 5051 & $-1.35^{+0.27}_{-0.59}$ & $-1.25$ &      5.17$\pm   1.02$ & 2.77$\pm 0.18$  \\
 HD64606 &   5318$^{+34}_{-33}$  & 5206 & $-0.28^{+0.04}_{-0.04}$ & $-0.93$ &     52.01$\pm   1.85$ & 4.61$\pm 0.08$  \\
 HD84937 &   6406$^{+192}_{-196}$  & 6357 & $-2.43^{+0.25}_{-0.31}$ & $-2.10$ &     12.44$\pm   1.06$ & 4.14$\pm 0.10$  \\
 HD94028 &   6059$^{+102}_{-118}$  & 6060 & $-1.70^{+0.13}_{-0.18}$ & $-1.38$ &     19.23$\pm   1.13$ & 4.36$\pm 0.09$  \\
 HR1083 &   6372$^{+31}_{-32}$  & 6695 & $-0.68^{+0.03}_{-0.03}$ & $-0.14$ &     46.65$\pm   0.48$ & 3.85$\pm 0.07$  \\
 HR1729 &   5817$^{+19}_{-16}$  & 5824 & $+0.24^{+0.02}_{-0.01}$ & $-0.04$ &     79.08$\pm   0.90$ & 4.20$\pm 0.07$  \\
 HR203 &   5751$^{+46}_{-56}$  & 5793 & $-0.18^{+0.05}_{-0.04}$ & $-0.25$ &     31.39$\pm   1.03$ & 3.98$\pm 0.07$  \\
 HR2721 &   5855$^{+18}_{-21}$  & 5913 & $-0.20^{+0.01}_{-0.03}$ & $-0.27$ &     59.31$\pm   0.69$ & 4.28$\pm 0.07$  \\
 HR3262 &   6159$^{+53}_{-35}$  & 6301 & $-0.54^{+0.09}_{-0.03}$ & $-0.26$ &     55.17$\pm   0.93$ & 4.17$\pm 0.07$  \\
 HR3538 &   5615$^{+16}_{-20}$  & 5687 & $+0.07^{+0.01}_{-0.02}$ & $+0.02$ &     58.50$\pm   0.88$ & 4.36$\pm 0.07$  \\
 HR3648 &   5876$^{+19}_{-18}$  & 5830 & $+0.33^{+0.00}_{-0.00}$ & $-0.06$ &     51.12$\pm   0.72$ & 4.04$\pm 0.07$  \\
 HR3775 &   5922$^{+25}_{-21}$  & 6296 & $-0.77^{+0.02}_{-0.01}$ & $-0.21$ &     74.15$\pm   0.74$ & 3.50$\pm 0.07$  \\
 HR4277 &   5763$^{+10}_{-64}$  & 5811 & $+0.08^{+0.00}_{-0.16}$ & $+0.00$ &     71.04$\pm   0.66$ & 4.22$\pm 0.07$  \\
 HR4421 &   6329$^{+52}_{-48}$  & 6623 & $-1.09^{+0.06}_{-0.06}$ & $-0.51$ &     30.40$\pm   0.60$ & 3.87$\pm 0.07$  \\
 HR4540 &   5942$^{+28}_{-17}$  & 6065 & $-0.08^{+0.03}_{-0.01}$ & $+0.10$ &     91.74$\pm   0.77$ & 3.94$\pm 0.07$  \\
 HR4657 &   6174$^{+51}_{-52}$  & 6267 & $-0.90^{+0.05}_{-0.06}$ & $-0.66$ &     44.34$\pm   1.01$ & 4.31$\pm 0.07$  \\
 HR4785 &   5758$^{+10}_{-16}$  & 5814 & $-0.25^{+0.01}_{-0.03}$ & $-0.19$ &    119.46$\pm   0.83$ & 4.33$\pm 0.07$  \\
 HR483 &   5775$^{+6}_{-11}$  & 5825 & $+0.07^{-0.02}_{-0.01}$ & $-0.04$ &     79.09$\pm   0.83$ & 4.28$\pm 0.07$  \\
 HR4845 &   5858$^{+21}_{-16}$  & 5868 & $-0.42^{+0.03}_{-0.01}$ & $-0.51$ &     57.57$\pm   0.64$ & 4.38$\pm 0.07$  \\
 HR5011 &   5804$^{+18}_{-22}$  & 5920 & $-0.12^{+0.00}_{-0.03}$ & $+0.10$ &     55.71$\pm   0.85$ & 4.10$\pm 0.07$  \\
 HR5235 &   5637$^{+17}_{-14}$  & 5943 & $-0.15^{+0.00}_{-0.00}$ & $+0.20$ &     88.17$\pm   0.75$ & 3.38$\pm 0.07$  \\
 HR5447 &   6622$^{+38}_{-38}$  & 6734 & $-0.75^{+0.04}_{-0.04}$ & $-0.41$ &     64.66$\pm   0.72$ & 4.18$\pm 0.07$  \\
 HR5868 &   5835$^{+18}_{-14}$  & 5847 & $+0.12^{+0.02}_{-0.01}$ & $-0.04$ &     85.08$\pm   0.80$ & 4.16$\pm 0.07$  \\
 HR5914 &   5815$^{+16}_{-15}$  & 5831 & $-0.44^{+0.02}_{-0.01}$ & $-0.46$ &     63.08$\pm   0.54$ & 3.99$\pm 0.07$  \\
 HR5933 &   6175$^{+20}_{-20}$  & 6268 & $-0.31^{+0.01}_{-0.02}$ & $-0.18$ &     89.92$\pm   0.72$ & 4.09$\pm 0.07$  \\
 HR5968 &   5755$^{+17}_{-22}$  & 5745 & $-0.13^{+0.02}_{-0.02}$ & $-0.22$ &     57.38$\pm   0.71$ & 4.17$\pm 0.07$  \\
 HR6243 &   6061$^{+78}_{-128}$  & 6361 & $-0.36^{+0.05}_{-0.22}$ & $-0.03$ &     27.04$\pm   1.08$ & 3.24$\pm 0.08$  \\
 HR6315 &   6157$^{+17}_{-19}$  & 6215 & $-0.33^{+0.01}_{-0.02}$ & $-0.18$ &     66.28$\pm   0.48$ & 4.22$\pm 0.07$  \\
 HR6458 &   5657$^{+12}_{-12}$  & 5633 & $-0.18^{+0.01}_{-0.01}$ & $-0.38$ &     69.48$\pm   0.56$ & 4.29$\pm 0.07$  \\
 HR6775 &   5941$^{+18}_{-18}$  & 6001 & $-0.76^{+0.02}_{-0.01}$ & $-0.54$ &     63.88$\pm   0.55$ & 4.15$\pm 0.07$  \\
 HR7061 &   5927$^{+30}_{-32}$  & 6301 & $-0.30^{+0.03}_{-0.03}$ & $-0.09$ &     52.37$\pm   0.68$ & 3.61$\pm 0.07$  \\
 HR7560 &   5920$^{+25}_{-25}$  & 6047 & $-0.06^{+0.01}_{-0.01}$ & $+0.03$ &     51.57$\pm   0.77$ & 4.04$\pm 0.07$  \\
 HR784 &   6118$^{+35}_{-32}$  & 6209 & $-0.10^{+0.02}_{-0.03}$ & $-0.01$ &     46.42$\pm   0.82$ & 4.26$\pm 0.07$  \\
 HR8181 &   5958$^{+15}_{-16}$  & 6244 & $-1.32^{+0.01}_{-0.03}$ & $-0.62$ &    108.50$\pm   0.59$ & 4.27$\pm 0.07$  \\
 HR8354 &   6119$^{+44}_{-41}$  & 6378 & $-1.07^{+0.05}_{-0.04}$ & $-0.59$ &     36.15$\pm   0.69$ & 3.84$\pm 0.07$  \\
 HR8665 &   5894$^{+26}_{-24}$  & 6184 & $-0.64^{+0.02}_{-0.01}$ & $-0.30$ &     61.54$\pm   0.77$ & 3.76$\pm 0.07$  \\
 HR8697 &   5956$^{+40}_{-40}$  & 6250 & $-0.64^{+0.02}_{-0.03}$ & $-0.23$ &     37.25$\pm   0.76$ & 3.72$\pm 0.07$  \\
 HR8729 &   5652$^{+17}_{-17}$  & 5669 & $+0.25^{+0.01}_{-0.01}$ & $+0.08$ &     65.10$\pm   0.76$ & 4.27$\pm 0.07$  \\
 HR8969 &   6078$^{+28}_{-24}$  & 6198 & $-0.24^{+0.03}_{-0.02}$ & $-0.17$ &     72.51$\pm   0.88$ & 4.00$\pm 0.07$  \\
\enddata
\end{deluxetable}

\end{document}